\documentclass[ammsymb, twocolumn, prb, superscriptaddress, floats, nobibnotes, a4paper]{revtex4-1}

\usepackage[T1]{fontenc}
\usepackage{bm}
\usepackage{graphicx}
\usepackage{natbib}
\usepackage{epstopdf}
\usepackage[euler]{textgreek}
\usepackage{hyperref}
\usepackage{color}

\usepackage[caption=false]{subfig}

\usepackage[labeled,resetlabels]{multibib}

\newcites{SI}{SupI}

\begin{document}
	\title{Optical response of monolayer, few-layer and bulk tungsten disulfide}
	
	\author{Maciej R. Molas}
	\email{maciej.molas@gmail.com}
	\affiliation{Laboratoire National des Champs Magn\'etiques Intenses, CNRS-UGA-UPS-INSA-EMFL, 25, avenue des Martyrs, 38042 Grenoble, France} 
	\author{Karol Nogajewski}
	\email{karol.nogajewski@lncmi.cnrs.fr}
	\affiliation{Laboratoire National des Champs Magn\'etiques Intenses, CNRS-UGA-UPS-INSA-EMFL, 25, avenue des Martyrs, 38042 Grenoble, France} 
	\author{Artur O. Slobodeniuk}
	\affiliation{Laboratoire National des Champs Magn\'etiques Intenses, CNRS-UGA-UPS-INSA-EMFL, 25, avenue des Martyrs, 38042 Grenoble, France} 
	\author{Johannes Binder}
	\affiliation{Laboratoire National des Champs Magn\'etiques Intenses, CNRS-UGA-UPS-INSA-EMFL, 25, avenue des Martyrs, 38042 Grenoble, France} 
	\affiliation{Faculty of Physics, University of Warsaw, ul. Pasteura 5, 02-093 Warszawa, Poland} 
	\author{Miroslav Bartos}
	\affiliation{Laboratoire National des Champs Magn\'etiques Intenses, CNRS-UGA-UPS-INSA-EMFL, 25, avenue des Martyrs, 38042 Grenoble, France} 
	\author{Marek Potemski}
	\email{marek.potemski@lncmi.cnrs.fr}
	\affiliation{Laboratoire National des Champs Magn\'etiques Intenses, CNRS-UGA-UPS-INSA-EMFL, 25, avenue des Martyrs, 38042 Grenoble, France} 
	\affiliation{Faculty of Physics, University of Warsaw, ul. Pasteura 5, 02-093 Warszawa, Poland}

\begin{abstract}
	
	We present a comprehensive optical study of thin films of tungsten disulfide (WS$_2$) with layer thicknesses ranging from mono- to octalayer and in the bulk limit. It is shown that the optical band-gap absorption of monolayer WS$_2$ is governed by competing resonances arising from one neutral and two distinct negatively charged excitons whose contributions to the overall absorption of light vary as a function of temperature and carrier concentration. The photoluminescence response of monolayer WS$_2$ is found to be largely dominated by disorder/impurity- and/or phonon-assisted recombination processes. The indirect band-gap luminescence in multilayer WS$_2$ turns out to be a phonon-mediated process whose energy evolution with the number of layers surprisingly follows a simple model of a two-dimensional confinement. The energy position of the direct band-gap response (A and B resonances) is only weakly dependent on the layer thickness, which underlines an approximate compensation of the effect of the reduction of the exciton binding energy by the shrinkage of the apparent band gap. The A-exciton absorption-type spectra in multilayer WS$_2$ display a non-trivial fine structure which results from the specific hybridization of the electronic states in the vicinity of the K-point of the Brillouin zone. The effects of temperature on the absorption-like and photoluminescence spectra of various WS$_2$ layers are also quantified.
	
\end{abstract}

\maketitle

%%%MAIN TEXT%%%%
\section{Introduction}\label{sec:intro}

Semiconducting transition metal dichalcogenides (S-TMDs) such as MoS$_2$, MoSe$_2$, WS$_2$, WSe$_2$, and MoTe$_2$ have recently attracted a considerable attention due to their unique electronic structures and resultant optical properties.\cite{novoselov2005,wang2012} 

When thinned down to the monolayer, S-TMDs transform from indirect- to direct-band gap, optically-bright semiconductors\cite{mak2010,splendiani2010,zhao2012} and exhibit a number of appealing optical phenomena related, in particular, to valley-selective circular dichroism,\cite{mak2012,xiao,cao,jones,lagarde}  and non-trivial effects of Coulomb interaction in the two-dimensional geometry.\cite{mak2013,ross,britnell,ramas,cheiw,qiu,chernikov,zhu,ye} Atomically thin S-TMDs may find their practical application in optoelectronics (light emitting and photo-diodes,\cite{Withers2015,Withers2015NN,Genevie2016,Palacio2016,schwarz2016,binder2017} concepts of valleytronic\cite{Yu2014,Zhang2014,Seyler2015,Lee2016,Schaibley2016} and flexible devices\cite{britnell,kazuma2015,amani2015,gao2017}) what together with the scientific curiosity to exploring the properties of new systems stimulates the pertinent efforts to understand the fundamental properties of these systems. Whereas many works have, so far, been focused on S-TMD monolayers, less attention has been paid on systematic investigations of the optical response of S-TMDs as a function of number of layers.\cite{arora,aroramose2}

In this paper, we present a thorough study of the optical 
properties of thin films of WS$_2$ with thicknesses ranging from
monolayer (1 ML) to octalayer (8 MLs) and of a 32 nm thick
bulk-like flake, carried out in a wide temperature range (5-300~K)
using micro-photoluminescence (\textmu-PL), micro-photoluminescence 
excitation (\textmu-PLE) and micro-reflectance
contrast (\textmu-RC) spectroscopy techniques. Research of this
type, which plays an important role for elaborating the electronic
band structure as well as for identifying the apparent excitonic resonances, has not
yet been done in such a broad scope in the case of WS$_2$. 
Apart of the Introduction and Conclusions parts, our paper is composed of three
(\ref{sec:monolayer}-\ref{sec:absorption}) main sections completed by the Methods' paragraph and the ESI. 
First, in Section \ref{sec:monolayer}, we focus on the characterization of the optical response of the monolayer WS$_2$
and discuss the nature of various excitonic resonances (due to neutral, charged, and localized electron-hole complexes) 
which determine the absorption-like and photoluminescence spectra of this system. 
The evolution of the PL spectra as a function of number of monolayers, $N$, is presented in Section \ref{sec:indirect} 
and serves us to demonstrate the phonon assisted character of indirect recombination processes in multilayer WS$_2$
and to reproduce the $N$-dependence of the energy of the PL band within a simple model of carriers confined in 
a rectangular potential well. Section \ref{sec:absorption} is dedicated to the analysis of the RC spectra of multilayer WS$_2$,
whose clear, multiple-resonance character at the direct bandgap edge is identified as due to a specific 
hybridization scheme of electronic states at the K points of the Brillouin zone of 2H-stacked S-TMD layers. 
The samples used for investigations and our experimental techniques are described in the Methods' section. 
Additional results are presented in the ESI, which include the theoretical approach to account for the 
hybridization of electronic states in multilayer WS$_2$ and the analysis of the effect of temperature on 
the energy positions of excitonic resonances in the investigated samples.

\begin{figure}[t]
	\centering
	\includegraphics[width=1\linewidth]{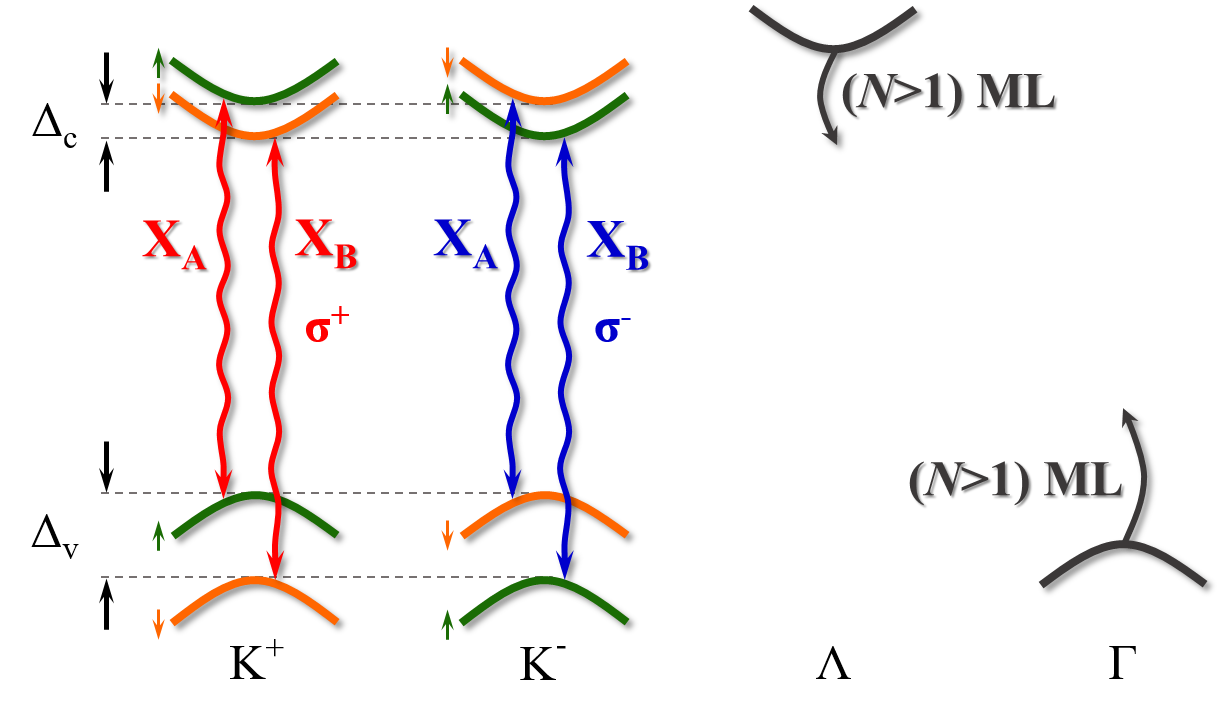}
	\caption{Diagram of relevant subbands in the conduction and valence bands at the K$^+$, K$^-$, $\Lambda$, and $\Gamma$ points of the Brillouin zone in monolayer WS$_2$. The green (orange) curves indicate the spin-up (spin-down) subbands. Drawn in grey are subbands without the spin projection. The red and blue wavy arrows show the X$_\textrm{A}$ and X$_\textrm{B}$ transitions active in the $\sigma^+$ and $\sigma^-$ polarizations, respectively. $\Delta_\textrm{c}$ and $\Delta_\textrm{v}$ denote the corresponding spin-orbit splitting in the conduction and valence bands. The grey arrows indicate the directions of movement of the band edges at the $\Lambda$ and $\Gamma$ points of the Brillouin zone, which in multilayer WS$_2$ progressively shift downwards and upwards, respectively, as the number of layers, $N$, is increased.}
	\label{fig:theory}
\end{figure}

Helpful for the presentation of our results is to highlight the essential facts about the near- band gap electronic structure of WS$_2$ mono- and multilayers.
A diagram of the electronic bands alignment in the
vicinity of the fundamental band gap of a WS$_2$ monolayer is shown in
Fig.~\ref{fig:theory}. Alike for the case of any S-TMD, a
monolayer WS$_2$ is a direct band-gap semiconductor with the minima
(maxima) of the conduction (valence) band located at the K$^+$ and
K$^-$ points of the Brillouin zone (BZ). A strong spin-orbit coupling, inherited
from heavy metal atoms, leads to spin-split and spin-polarized
subbands in both the valence band (VB) and the conduction band (CB). The amplitude of the spin-orbit splitting at the top of the valence band (electronic states
predominantly built of $d_{\pm2}$ orbitals of the metal atoms) is large as it reaches
$\sim400$~meV.\cite{ye,chernikov,zhao2012,kozawa,zeng,zhu,he} The
spin-orbit splitting for the bottom conduction band states (mostly built
of $d_{0}$ orbitals of the metal atoms) is considered to be significantly
smaller and as such it was largely neglected in the initial works
on S-TMD monolayers. More recently, however, its magnitude has been estimated to amount to $\sim$30
meV in monolayer WS$_{2}$,\cite{liu,kormanyos} based on the band structure calculations
including contributions of the opposite sign coming from second order
effects for $d_{0}$ metal orbitals and an additional input from the
admixed $p$-orbitals of the chalcogen atoms. Importantly, the expected
alignment of the CB spin-orbit-split subbands in monolayer
WS$_{2}$ is such that the transition between the uppermost VB
subband and the lowest CB subband is optically inactive.
Associated with this transition, the lowest energy component of
the A-exciton is therefore optically dark whereas the optically
active component of the A-exciton is linked to the upper CB
subband and appears at a higher energy. A similar alignment of the
spin-orbit-split subbands is expected for monolayer WSe$_{2}$ (and
likely MoS$_{2}$).\cite{molas} It contrasts with MoSe$_{2}$ and MoTe$_{2}$
monolayers which exhibit the lowest energy component of the A-exciton
that is optically bright.\cite{molas}

While monolayers of S-TMDs have been investigated to date in many works, the
few-layer S-TMDs have attracted so far less attention and their
properties, including those of WS$_2$ multilayers, are not well
known. What is generally accepted is that all $N$-ML S-TMDs are
indirect semiconductors for $N>1$ (perhaps with the exception of
still direct band-gap bilayer MoTe$_{2}$~\cite{lezama2015,froehlicher2016,robert2016}). This is understood in
terms of a significant evolution of the band edges at the
$\Lambda$\cite{zhao2013,gutierez} and $\Gamma$ points of the BZ: a downward
shift of the local CB minimum at the $\Lambda$ point and an upward
shift of the local VB maximum at the $\Gamma$ point, see Fig.~\ref{fig:theory}. Although the
indirect band gap in S-TMD multilayers is likely between the
$\Lambda$-point minimum of the CB and the $\Gamma$-point maximum of the VB, a more detailed
description of the electronic bands (and the optical response) in these
structures is definitely less obvious.

\section{Near band-edge excitonic resonances in monolayer WS$_2$}\label{sec:monolayer}

\begin{center}
	\begin{figure}[t]
		\centering
		\includegraphics[width=1\linewidth]{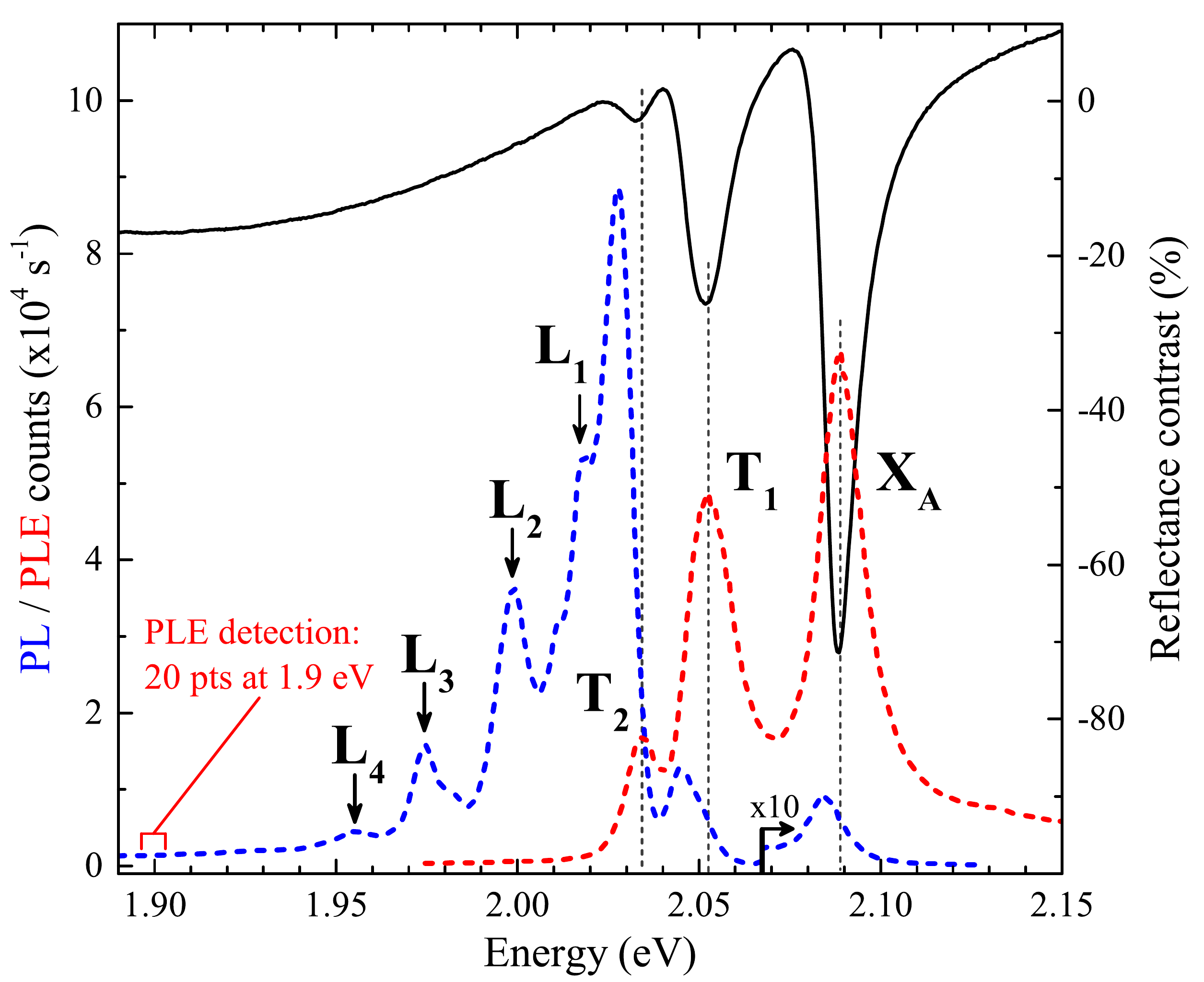}%
		\caption{Comparison of the photoluminescence (PL, blue dashed curve), photoluminescence excitation (PLE, red dashed curve), and reflectance contrast (RC, black solid curve) spectra measured on the WS$_2$ monolayer at $T$=5~K.}
		\label{fig:PL_PLE_RC_ML}
	\end{figure}
\end{center}

In this section we focus on fundamental optical transitions in monolayer WS$_2$, $i.e.$, on those transitions which appear in the vicinity of the optical band gap (A-exciton resonance). Figure~\ref{fig:PL_PLE_RC_ML} illustrates the optical-band-gap response of our WS$_2$ monolayers as investigated with \textmu-PL, \textmu-PLE and \textmu-RC measurements. The RC and PLE spectra show up to three absorption-type resonances, denoted in Fig.~\ref{fig:PL_PLE_RC_ML} as X$_\mathrm{A}$, T$_1$, and T$_2$. These resonances appear in the RC spectra in the form of dips, whose energy positions coincide with the central energies of peaks observed in the PLE spectra. The X$_\mathrm{A}$, T$_1$, and T$_2$ transitions are also visible in the emission spectra though the corresponding PL peaks are slightly ($\sim$5~meV) redshifted with respect to their absorption counterparts (Stokes shift).

\begin{center}
	\begin{figure}[t]
		\centering
		\includegraphics[width=0.8\linewidth]{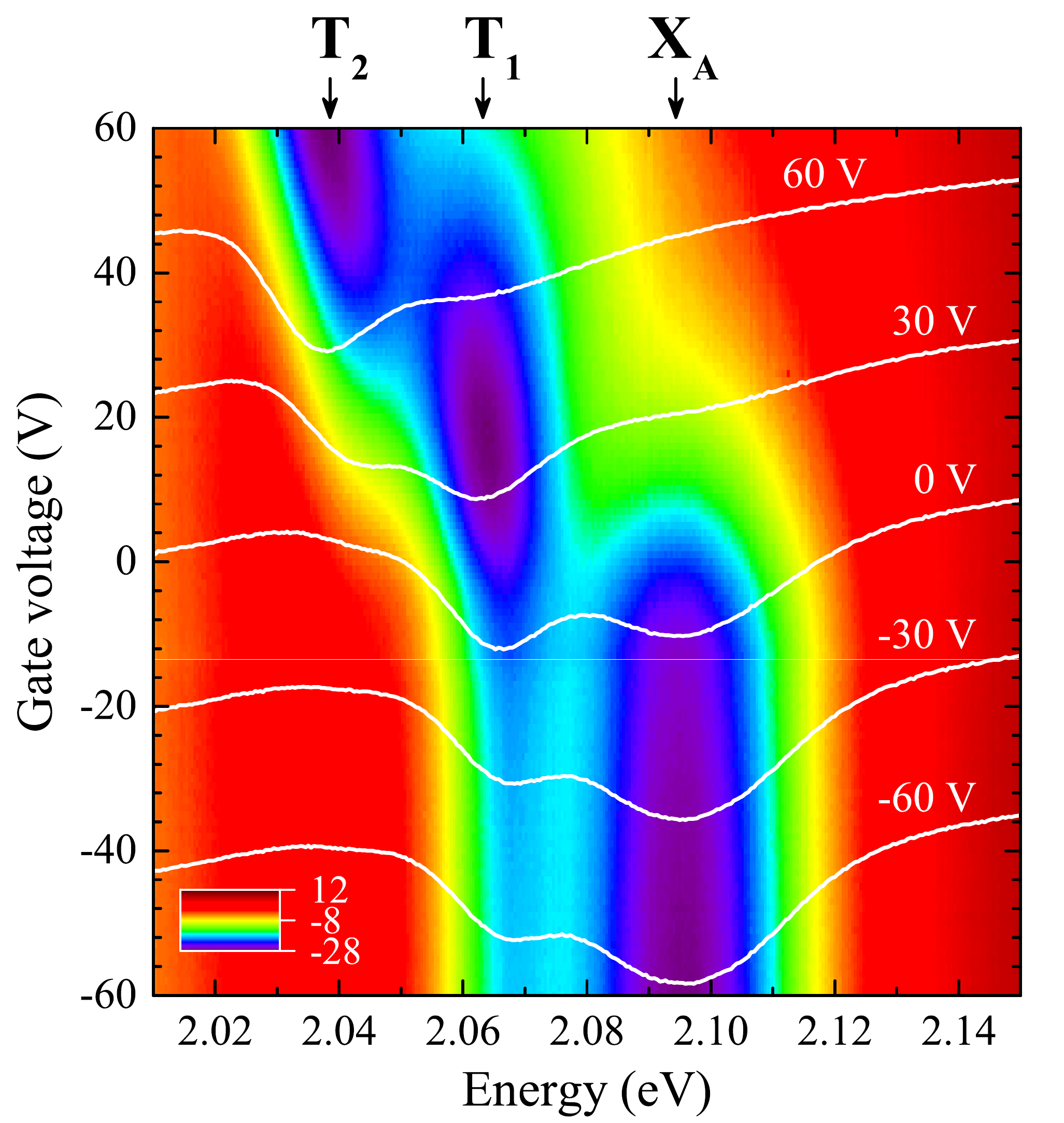}%
		\caption{False-colour map of the reflectance contrast (RC) spectra measured on the WS$_2$ monolayer at $T$=5~K as a function of gate voltage ranging from -60~V up to 60~V. White curves superimposed on the map represent the RC spectra recorded at selected gate voltages.}
		\label{fig:RC_Gatea}
	\end{figure}
\end{center}

\begin{figure}[h]
	\centering
	\includegraphics[width=0.75\linewidth]{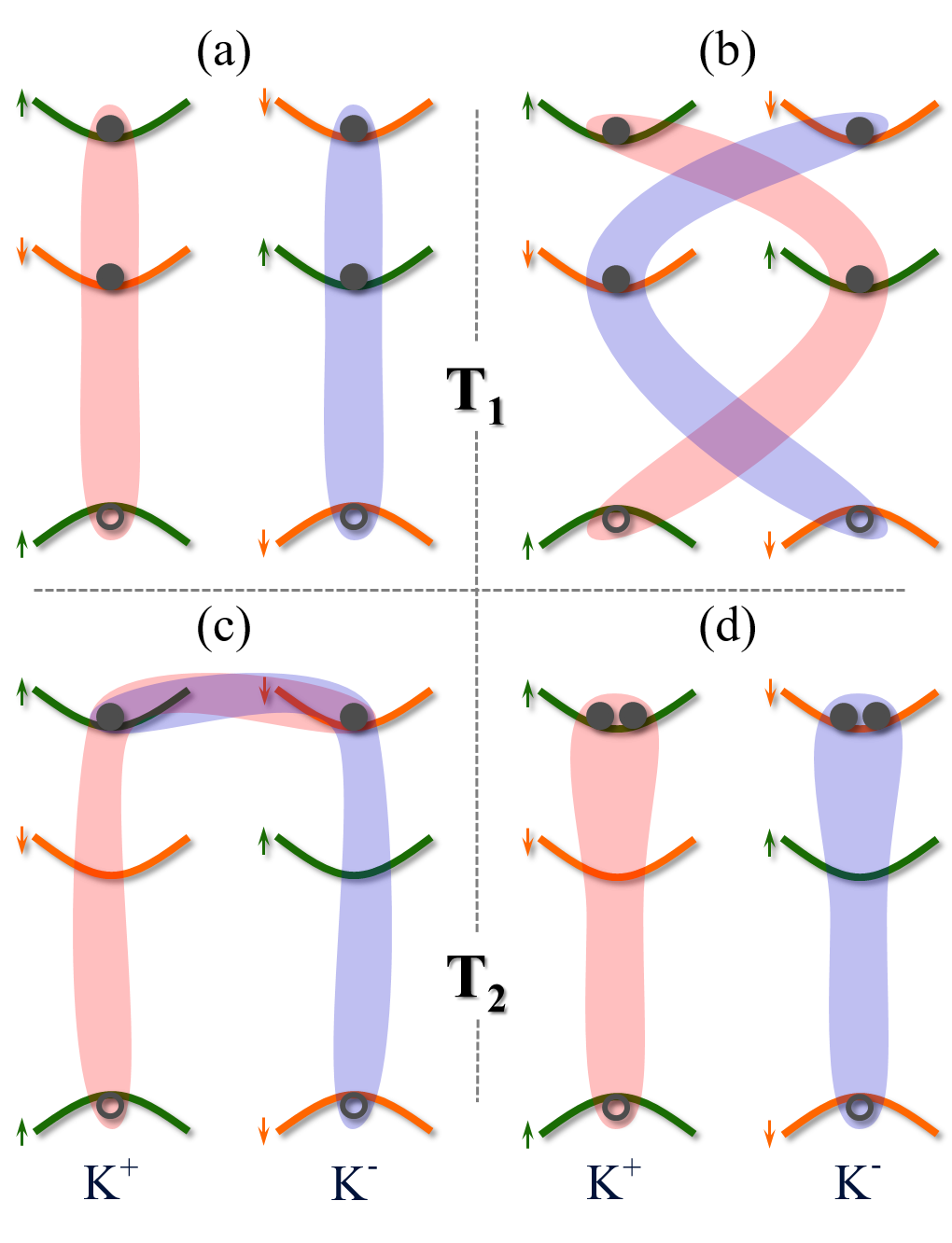}
	\caption{Schematic illustration of possible configurations for the charge carriers constituting bright negatively charged excitonic complexes (trions) T$_1$ and T$_2$ in a WS$_2$ monolayer at the K$^\pm$ points of the Brillouin zone. The (a) and (c) panels present the spin-singlet configurations, while the (b) and (d) panels show the spin-triplet ones. The green (orange) curves indicate the spin-up (spin-down) subbands. The electrons (holes) in the conduction (valence) band are represented by closed (open) gray circles. The red and blue shapes denote the constituent complexes in each doubly degenerate pair of trions.}
	\label{fig:trions}
\end{figure}

Whereas it is quite common that the optical-band-gap absorption 
of S-TMD monolayers displays two resonances due to neutral and 
charged excitons, the appearance of three absorption resonances
in our WS$_2$ monolayers is somehow striking. The assignment of the 
X$_\mathrm{A}$ resonance as arising from the neutral A-exciton 
is straightforward and stays in accordance with many other works on WS$_2$ 
monolayers.\cite{chernikov,scrace,chernikov2015} The origin of
T$_1$ and T$_2$ features is less clear. A possibility that T$_1$
is due to charged exciton\cite{chernikov,scrace,chernikov2015}
while T$_2$ is due to an impurity bound exciton has been recently 
put forward.\cite{jadczakWS2} To gain more insight into the character
of T$_1$ and T$_2$ transitions we present in Fig.~\ref{fig:RC_Gatea} 
the results of RC measurements performed on a back-gated WS$_2$ monolayer structure. 
The application of a positive gate voltage ($V$) in this device leads 
to an effective increase of the electron concentration ($n$) in 
the WS$_2$ monolayer. The absorption resonances observed in our 
gated device are somewhat broader than those measured on ungated 
monolayer flakes (see Fig.~\ref{fig:PL_PLE_RC_ML}) but the 
characteristic three,  X$_\mathrm{A}$, T$_1$, and T$_2$ features 
can well be recognized. In the regime of relatively low electron
concentration (negative gate voltage) the RC spectra show mostly 
the strong X$_\mathrm{A}$ and somewhat weaker T$_1$ resonance. 
With the increase of $n$ (the increase of positive $V$) 
the X$_\mathrm{A}$ resonance progressively weakens and eventually disappears from the spectra
measured at highest electron concentrations (60~V). The T$_1$ 
resonance is rather weakly affected by the gate voltage. Initially,
it slightly gains in intensity and remains apparent in the 
spectra for any $V$, although in a somewhat broadened form in the limit of 
high electron densities. In contrast, the T$_2$ resonance becomes clearly stronger with $V$ and finally dominates the RC spectra measured in the limit of high electron 
concentration. The data presented in Fig.~\ref{fig:RC_Gatea} are 
hardly conceivable with the assignment of the feature T$_2$ as 
due to the bound exciton resonance.\cite{jadczakWS2} The observed 
transfer of the oscillator strength as a function of electron density, 
progressively from the neutral exciton X$_\mathrm{A}$ to the T$_1$ 
and eventually to the T$_2$ resonance indicates that both T$_1$ 
and T$_2$ features can be associated with negatively charged excitons (X$^-$), 
though implying different configurations of these three-particle
complexes. As schematically shown in Fig.~\ref{fig:trions}, four pairs of doubly degenerate configurations for the bright X$^-$ complex are possible in a WS$_2$ monolayer. The formation of X$^-$ states 
in the configurations shown in the upper panels ((a) and (b)) of 
Fig.~\ref{fig:trions} is possible already at low electron densities 
(access electrons occupy the bottom CB subband). On the other hand, 
the X$^-$ states in the configurations shown in Fig.~\ref{fig:trions}(c)
and (d) are favoured at higher densities, $i.e.$, when access electrons
occupy also the upper CB subband. We speculate
that the T$_1$ resonance is associated with the X$^-$ states which 
involve the access electron in the bottom CB subband (Fig.~\ref{fig:trions}(c) 
and (d)) whereas the T$_2$ resonance corresponds to the X$^-$ states
involving the access electron from the upper CB subband (configurations
illustrated in bottom panels of Fig.~\ref{fig:trions}). Each T$_1$ and T$_2$
resonance comprises a pair of singlet (left panels of Fig.~\ref{fig:trions})
and triplet (right panels of Fig.~\ref{fig:trions}) states which can be 
further split in energy\cite{plechingerTRION} but not resolved within 
the linewidth of T$_1$ and T$_2$ features in the RC spectra measured.
It should be noted that our interpretation of T$_1$ and T$_2$ resonances
implies a significantly bigger binding energy for the charged exciton which involves
a pair of electrons from the same subband in the CB as compared to that 
involving a pair of electrons from different CB subbands. This conclusion 
awaits the confirmation in future theoretical study.

\begin{center}
	\begin{figure}[t]
		\centering
		\includegraphics[width=0.8\linewidth]{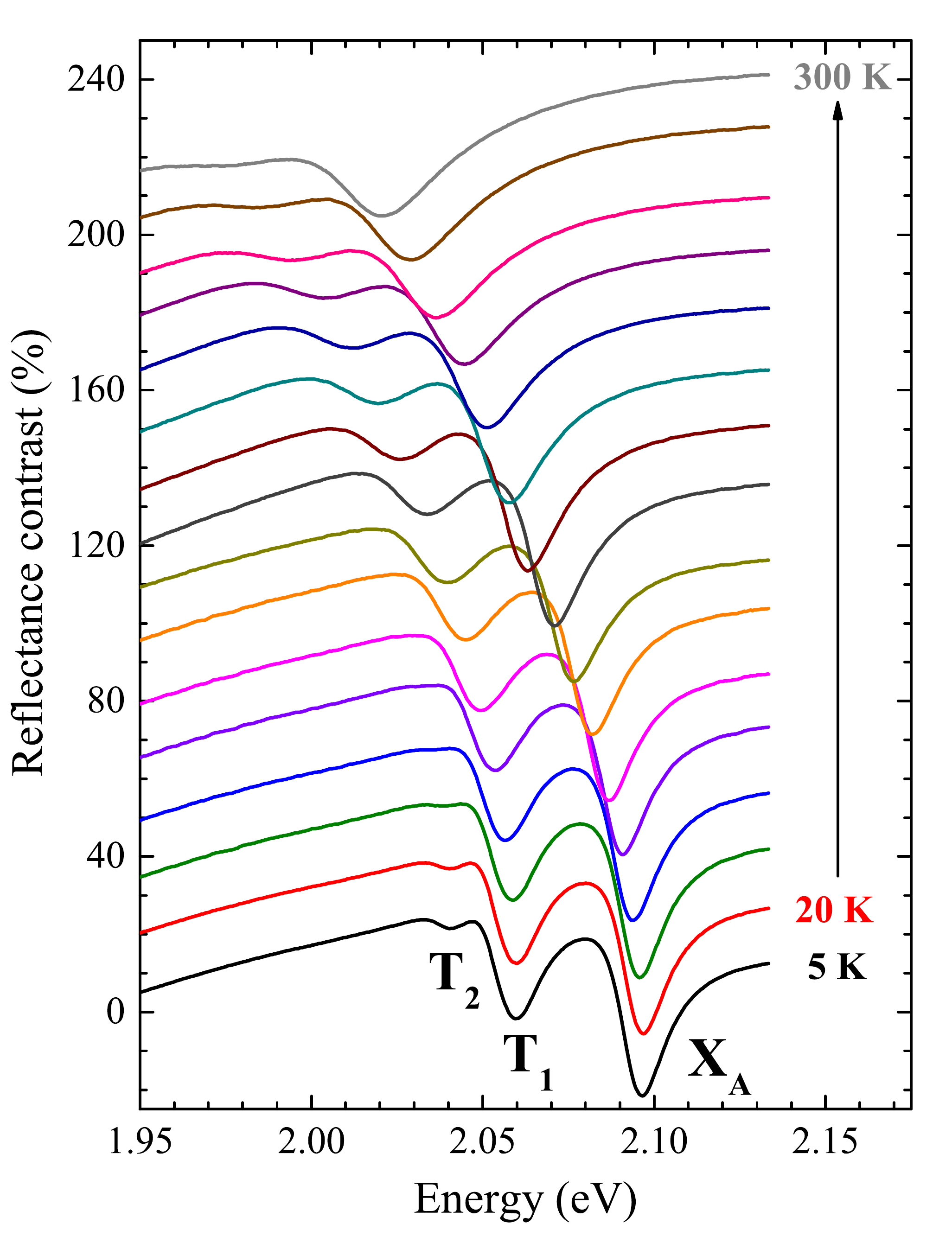}%
		\caption{Reflectance contrast spectra of monolayer WS$_2$ measured at different temperatures ranging from 5~K up to 300~K. The spectra are vertically shifted for clarity.}
		\label{fig:RC_T_ML}
	\end{figure}
\end{center}

In addition to carrier concentration, temperature is another parameter
known to affect the relative absorption strength of neutral and charged excitons.
Upon increase of temperature, neutral excitons often gain in intensity. 
This happens at the expense of an effective quenching of charge exciton resonances due to
weakening of their oscillator strength when they are associated with electrons 
occupying states of higher k-vectors.\cite{esser_Trion} The RC spectra measured as a function
of temperature are presented in Fig.~\ref{fig:RC_T_ML}. They were obtained for the 
WS$_2$ monolayer whose low-temperature spectrum has already been shown in Fig.~\ref{fig:PL_PLE_RC_ML}.
For this monolayer, all three X$_\mathrm{A}$, T$_1$, and T$_2$ resonances are present
in the RC spectrum measured at low temperature, though the T$_2$ feature is rather weakly
pronounced. We expect the electron concentration in this sample to be relatively low but enough
to set the Fermi energy slightly above the edge of the upper CB subband. Under such 
conditions, an increase of temperature may largely affect the k-space distribution 
of electrons in the upper CB subband, but should not significantly influence the distribution of electrons
in the bottom CB subband. This reasoning is in qualitative agreement with the observed 
rapid temperature quenching of the T$_2$ resonance accompanied by a survival of the T$_1$ 
resonance even up to room temperature. A detailed analysis of the RC spectra shown in Fig.~\ref{fig:RC_T_ML}, performed within the framework of transfer matrix method combined with the Lorentz oscillator model to determine the temperature evolution of the energy positions and linewidths of all three resonances (X$_\mathrm{A}$, T$_1$, and T$_2$) is presented in the ESI.

\begin{center}
	\begin{figure}[t]
		\centering
		\includegraphics[width=1\linewidth]{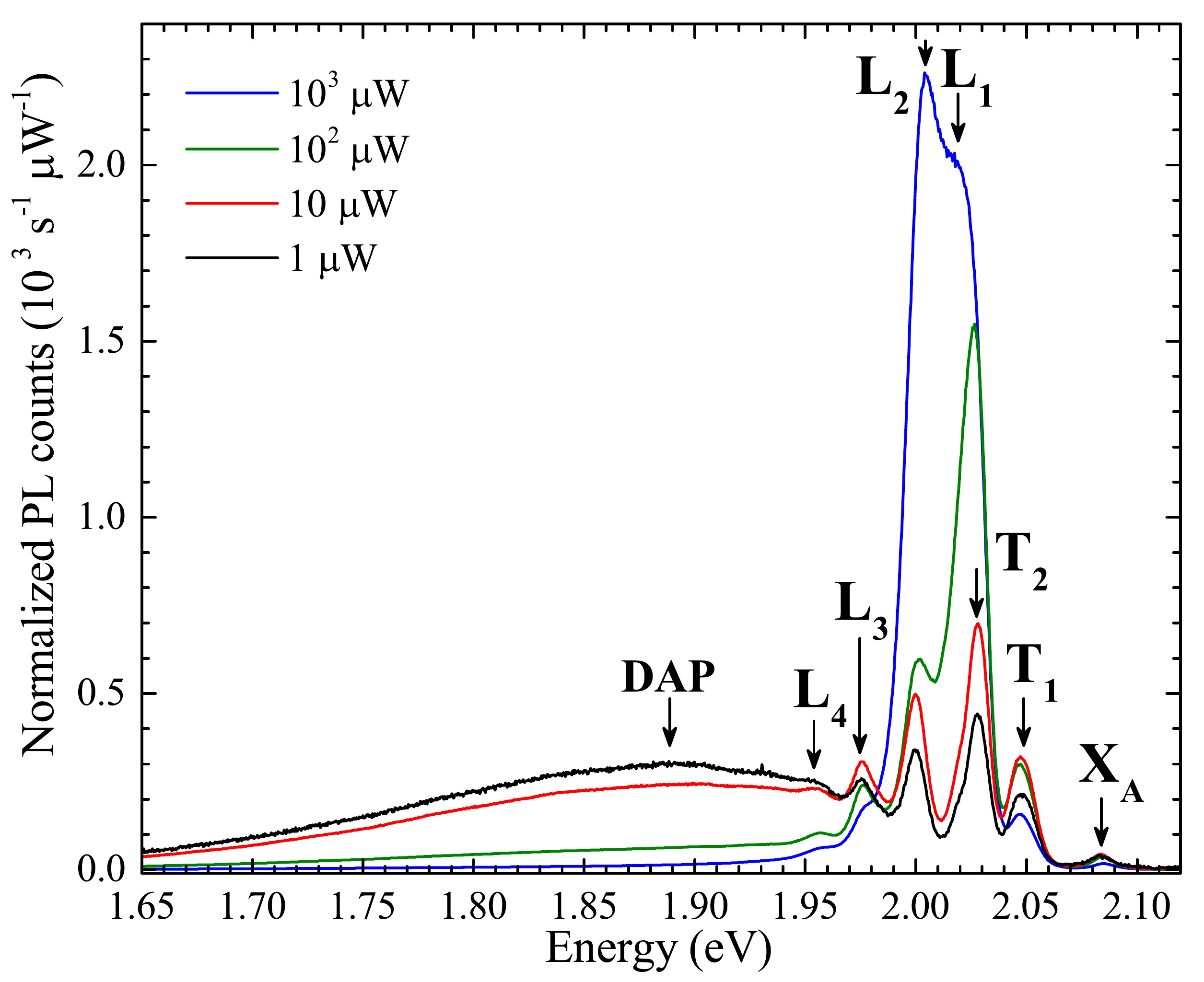}%
		\caption{Power dependence of low-temperature photoluminescence (PL) spectra measured on monolayer WS$_2$ with 2.41 eV laser light excitation. The intensities of the PL spectra are normalized by the excitation power. Under such normalization the intensity of every feature varying linearly with the excitation power should remain constant for all the spectra shown.}
		\label{fig:PL_Power}
	\end{figure}
\end{center}

\begin{center}
	\begin{figure}[t]
		\centering
		\includegraphics[width=0.8\linewidth]{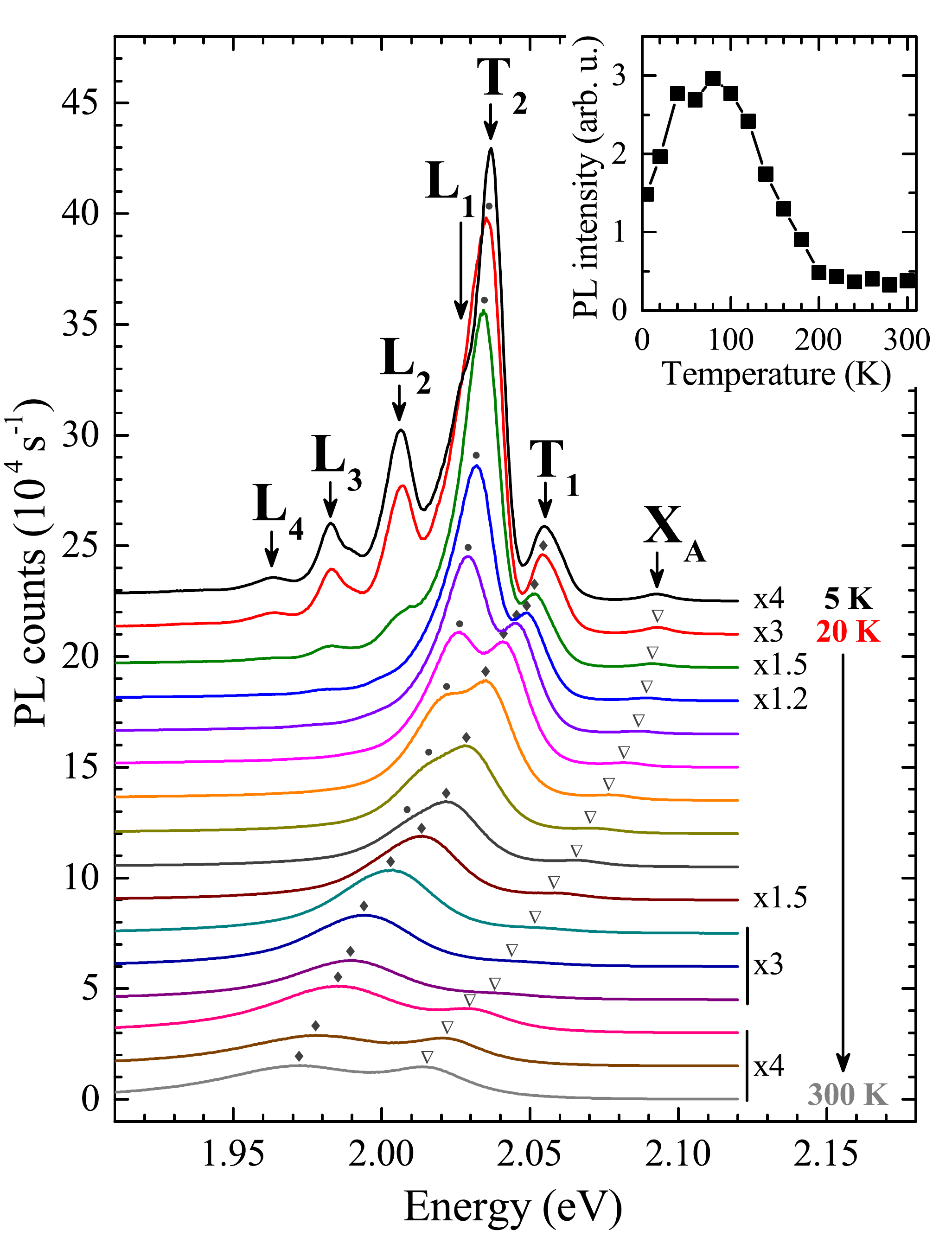}%
		\caption{Temperature evolution of photoluminescence spectra measured on monolayer WS$_2$ with 2.41 eV laser light excitation. The spectra are vertically shifted for clarity and some of them are multiplied by scaling factors in order to avoid their intersections with the neighbouring experimental curves or to make them better visible. The inset presents the integrated intensity of the whole PL spectrum as a function of temperature.}
		\label{fig:PL_T_ML}
	\end{figure}
\end{center}

Having discussed the absorption-type optical response of monolayer WS$_2$ let us now 
focus on the PL spectra (Figs \ref{fig:PL_Power} and \ref{fig:PL_T_ML}). 
One immediately notices their rather 
complex character which is particularly pronounced when these spectra are 
measured at low temperatures and display a number of eye-catching peaks/bands 
on the low energy side of the emission due to neutral and charged excitons (see 
features labelled as $\mathrm{L}_1,\ldots,\mathrm{L}_4$ and DAP in Figs. \ref{fig:PL_PLE_RC_ML}, 
\ref{fig:PL_Power}, and \ref{fig:PL_T_ML}). This behaviour is quite typical 
for semiconductors whose low-temperature PL spectra are often dominated 
by below-exciton emission bands associated with recombination processes 
assisted by disorder, defects/impurities and phonons. Such recombination 
channels ($e.g.$, due to bound/localized excitons, donor-acceptor 
recombination, phonon-assisted recombination) can be particularly pronounced
in monolayer WS$_2$ in which the exciton ground state is optically dark.\cite{kormanyos,molas}
Similar, multiple-peak low-temperature PL spectra are seen in monolayer WSe$_2$,\cite{arora} 
also characterized by the ground exciton state being dark. In contrast, 
monolayer MoSe$_2$, with the bright exciton ground state, shows often much
simpler PL spectra\cite{aroramose2} that even at low temperatures are dominated
by peaks due to neutral and charged excitons.

To discuss the origin of pronounced $\mathrm{L}_1,\ldots,\mathrm{L}_4$ and DAP features 
observed in the low-temperature PL spectra of our monolayer WS$_2$ we focus on
evolution of these spectra with the excitation power (see Fig.~\ref{fig:PL_Power}). 
The broad emission band, denoted as DAP in Fig.~\ref{fig:PL_Power}, is well pronounced in the 
limit of low excitation powers but clearly saturates when the laser power is 
increased. This observation is characteristic of the recombination spectrum 
of (dense) donor-acceptor pairs (DAP), which indeed frequently dominates the 
low-temperature weakly-excited photoluminescence in compensated semiconductors.\cite{bogardus,yu,singh,guo}
The $\mathrm{L}_1,\ldots,\mathrm{L}_4$ peaks, visible in emission but not in the RC (absorption-type) spectra, are commonly reported in the studies of WSe$_2$\cite{arora} 
and WS$_2$\cite{plechinger,shang} monolayers and loosely assigned to the 
recombination of "localized" excitons. With such an assignment, our observation
of prominent superlinear dependence of the intensity of the 
L$_1$ and L$_2$ emission peaks on the excitation power might be striking at the first sight. This superlinear
effect may be indicative of the biexciton recombination,\cite{You2015,plechinger,shang} though we hardly accept 
a possibility of the biexciton formation in our, after all not defect-free 
and electron-doped system. Moreover, time-resolved studies of analogous WS$_2$ monolayers 
show rather slow decay of the L$_1$ and L$_2$ luminescence\cite{smolenski_WS2} 
in contrast to what would be expected for the biexciton recombination. The 
superlinear dependence of the PL intensity on the excitation power may also 
be expected for the emission lines associated with recombination of impurity 
bound excitons.\cite{taguchi_1975,schmidt_1992,mostaani} A possible scenario, adequate for our n-type samples, is that 
the L$_1$ and L$_2$ peaks correspond to the recombination processes of excitons
bound to neutral acceptors (A$^0$X). The appearance of A$^0$X recombination is 
possible in n-type compensated semiconductors since the initially ionized 
acceptors (A$^-$) can trap the photoexcited holes thus becoming neutral
(A$^0$) and capable to bind an additional electron-hole pair (A$^0$X). The A$^0$X 
recombination appears to be a nonlinear process in compensated n-type semiconductors
what can account for the superlinear rise of the L$_1$ and L$_2$ intensities with the 
excitation power. The dynamics of optical recombination might be quite complex in
our samples and other nonlinear processes could be invoked as well, $e.g.$, the D$^+$X 
recombination of excitons bound to ionized donors (D$^+$) which can be activated after 
the DAP recombination processes. Instead, the D$^0$X and A$^-$X emission, naturally expected
in our n-type samples, could account for the PL peaks (L$_3$, L$_4$) whose intensities
scale more linearly with the excitation power. Finally, it is also logical to think 
that the below-exciton emission in monolayer WS$_2$ includes the recombination processes
of dark excitons, and in particular the phonon-assisted PL transitions of intervalley 
excitons. Such transitions are expected to display large Zeeman-type effects and 
the magneto-PL studies of monolayer WS$_2$ could help with their identification.

The PL spectra measured at moderate excitation power as a function of temperature are 
shown in Fig.~\ref{fig:PL_T_ML}. With increasing temperature, the consecutive low energy $\mathrm{L}_4,\ldots,\mathrm{L}_1$ peaks progressively disappear one after another from the spectrum. Three emission peaks (X$_\textrm{A}$, T$_2$, and T$_1$) are visible at intermediate temperatures, whereas only the neutral (X$_\textrm{A}$) and, notably, 
the charged exciton T$_1$ contribute to the PL spectrum at room temperature. X$_\textrm{A}$ as well as T$_1$
and T$_2$ resonances, even though not related to the ground exciton states in monolayer WS$_2$, are still 
seen in the PL spectra at low temperatures. These resonances are characterized by huge oscillator
strengths and/or extremely short radiative lifetimes in contrast to "localized" excitons whose 
PL decays much slower.\cite{smolenski_WS2} Thus, even if only weakly populated, the X$_\textrm{A}$, T$_1$, and T$_2$ peaks are observed in low-temperature PL and rapidly overcome the $\mathrm{L}_1,\ldots,\mathrm{L}_4$ emission when their population is increased at higher temperatures. As in the case of RC resonances, the effective 
thermal spreading of electrons in the upper CB subband accounts for the observed weakening of the
T$_2$ emission at highest temperatures. Worth noting is also the proposed unusual scheme for 
thermal ionization of bound versus charged excitons,\cite{ganchev}7 which, in accordance with our observations, 
implies a faster temperature quenching of bound excitons than of the charged excitons.

Temperature dependence of the integrated PL intensity, shown as an inset of Fig.~\ref{fig:PL_T_ML}, 
is overall governed by the interplay between different recombination channels including 
the non-radiative ones. The latter processes usually dominate at high temperatures what
results in PL quenching of about one order of magnitude when the temperature is changed from 5 to 300~K. 
Nevertheless, a clear, initial rise of the PL intensity in the range of 5-100~K should be noticed 
as well. This, in reference to a similar observation for monolayer WSe$_2$,\cite{arora,Wang2015,Zhang2015} fingerprints the semiconductor
with the lowest energy exciton being optically dark and thus displaying an effective thermal 
activation of the emission rising from higher energy bright excitons. The effect of temperature 
on the energy position (red shifts) of the PL peaks, well identifiable for the X$_\textrm{A}$ and T$_1$ excitons,
is similar to what is observed in the RC spectra (see the ESI for details).

\section{Indirect band-gap luminescence in WS$_2$ multilayers}\label{sec:indirect}

\begin{figure}[t]
	\centering
	\includegraphics[width=0.9\linewidth]{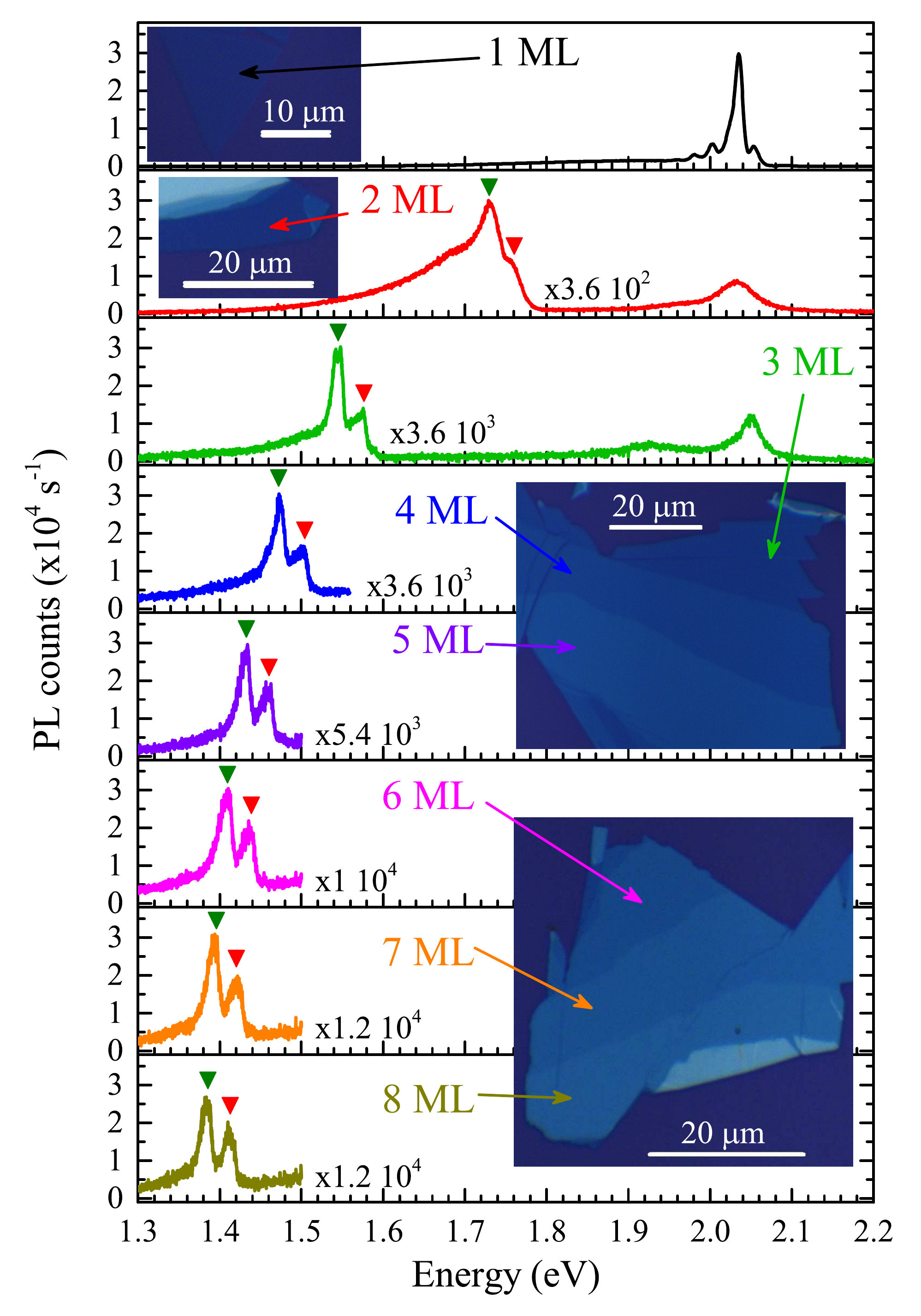}
	\caption{Photoluminescence spectra of thin films of WS$_2$ with thickness ranging from 1~ML to 8~MLs, measured at $T$=~5~K with the use of 2.41~eV laser light excitation. The red and green triangles denote the energy positions of the lines referred to in the text as I$_1$ and I$_2$. The insets show optical images of the flakes under study.}
	\label{fig:PL_vs_Thick}
\end{figure}

The PL spectra of moderately large-area WS$_{2}$ flakes composed of 1~ML up
to 8~MLs are presented in Fig.~\ref{fig:PL_vs_Thick}
(no measurable PL signal could be extracted from our bulk-like
flake of 32 nm thickness). As expected, the 1 ML-thick flake emits light
much more efficiently as compared to thicker flakes: the emission
intensity drops down by 4 orders of magnitude when the flake
thickness is increased from 1~ML to 8~MLs. This is a
characteristic feature of the whole family of S-TMDs, a very
first fingerprint of the change of the band edge alignment from a
direct band-gap monolayer to indirect band-gap
multilayers.\cite{mak2010,splendiani2010,gutierez}

\begin{center}
	\begin{figure}[t]
		\centering
		\includegraphics[width=0.6\linewidth]{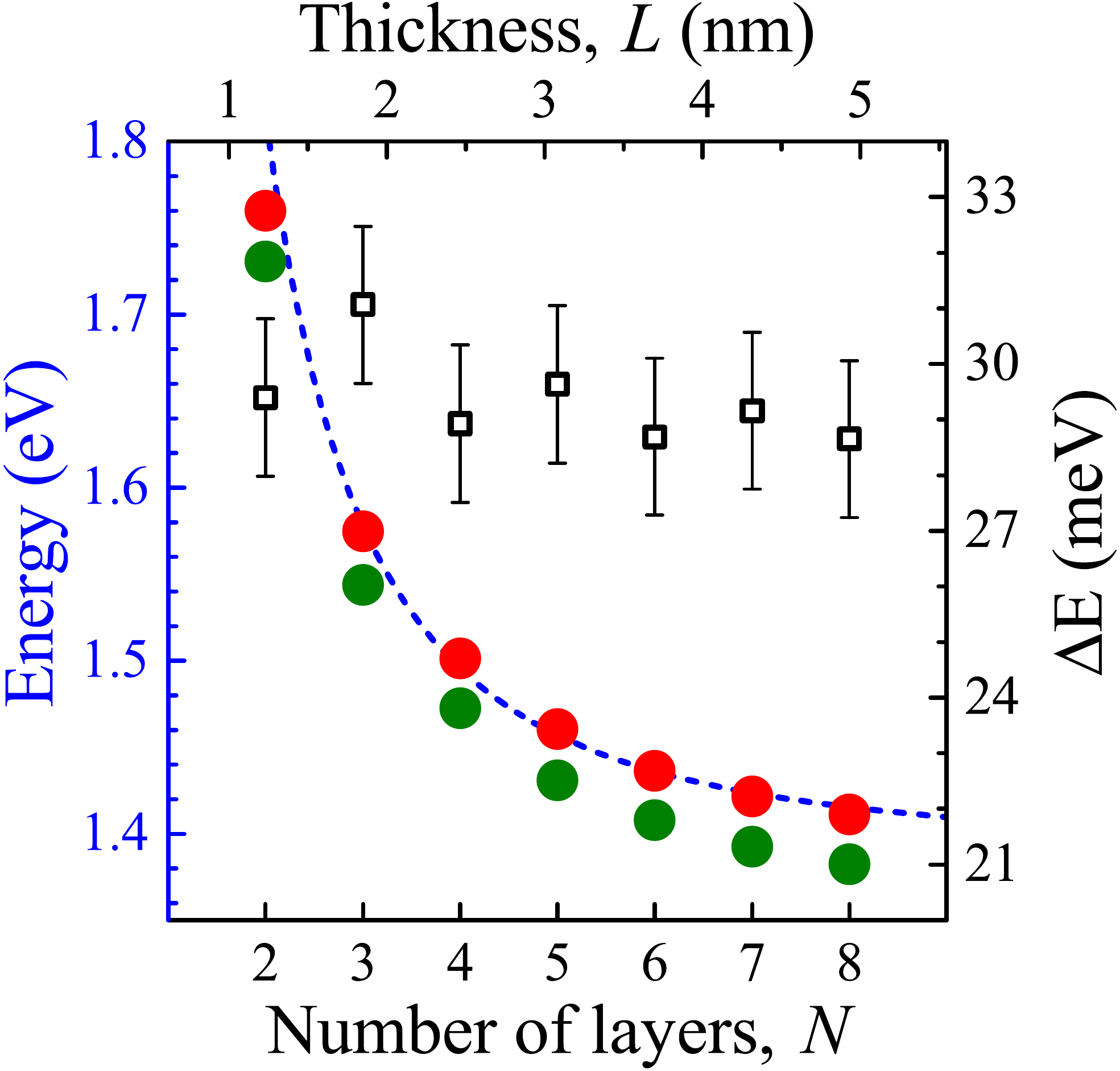}%
		\caption{Thickness dependence of energy positions of two well-resolved peaks, I$_1$ and I$_2$, constituting the emission band originating from indirect recombination processes (see Fig.~\ref{fig:PL_vs_Thick}) as well as the evolution of the energy separation ($\Delta$E) between them. The closed and open points represent the experimental results while the dashed curve is a fit to the data obtained with the aid of Eq. \ref{eq:qw}. }
		\label{fig:EnInd_vs_thick}
	\end{figure}
\end{center}

\begin{center}
	\begin{figure*}[t]
		\subfloat{}%
		\centering
		\includegraphics[width=0.4\linewidth]{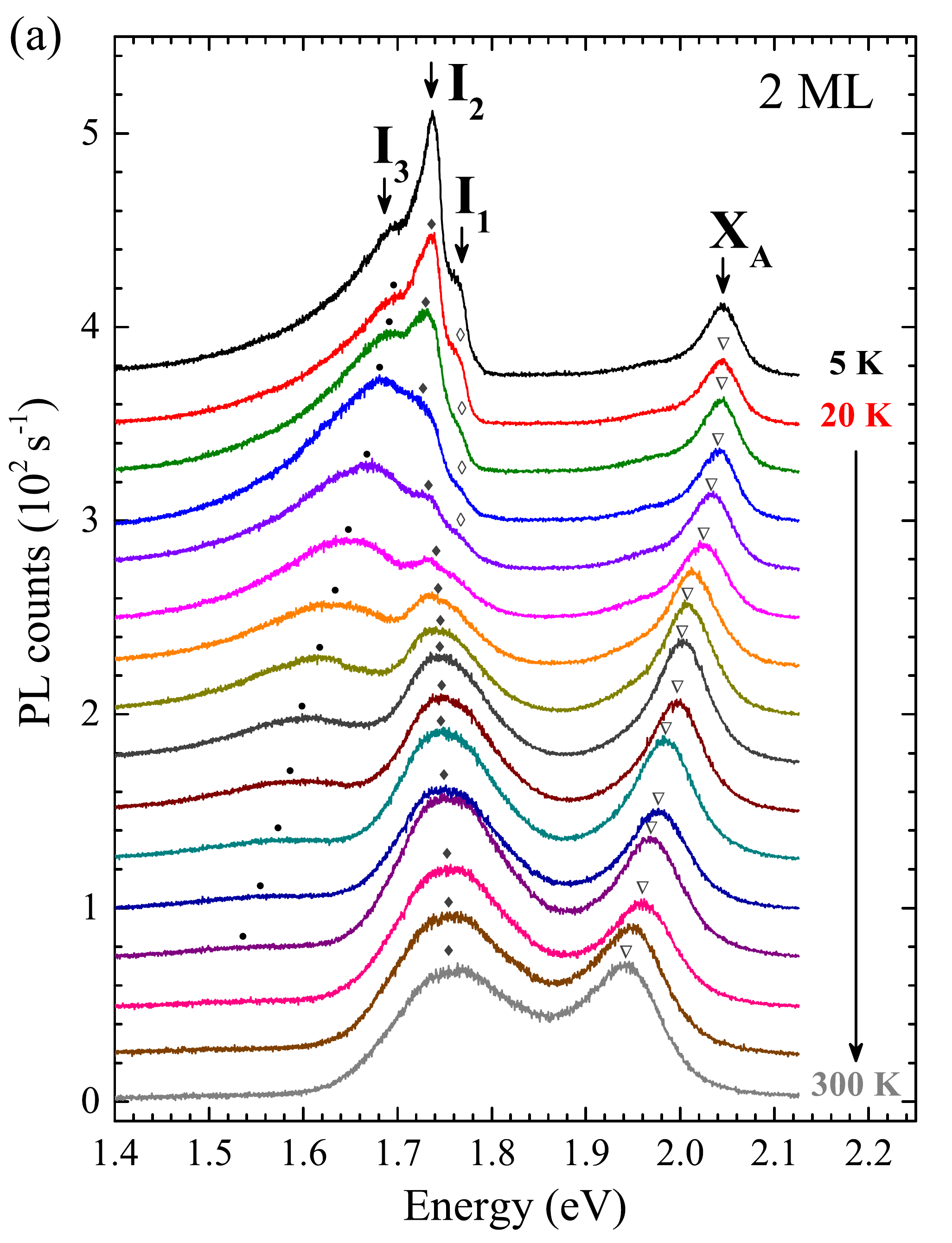}%
		\subfloat{}%
		\centering
		\includegraphics[width=0.4\linewidth]{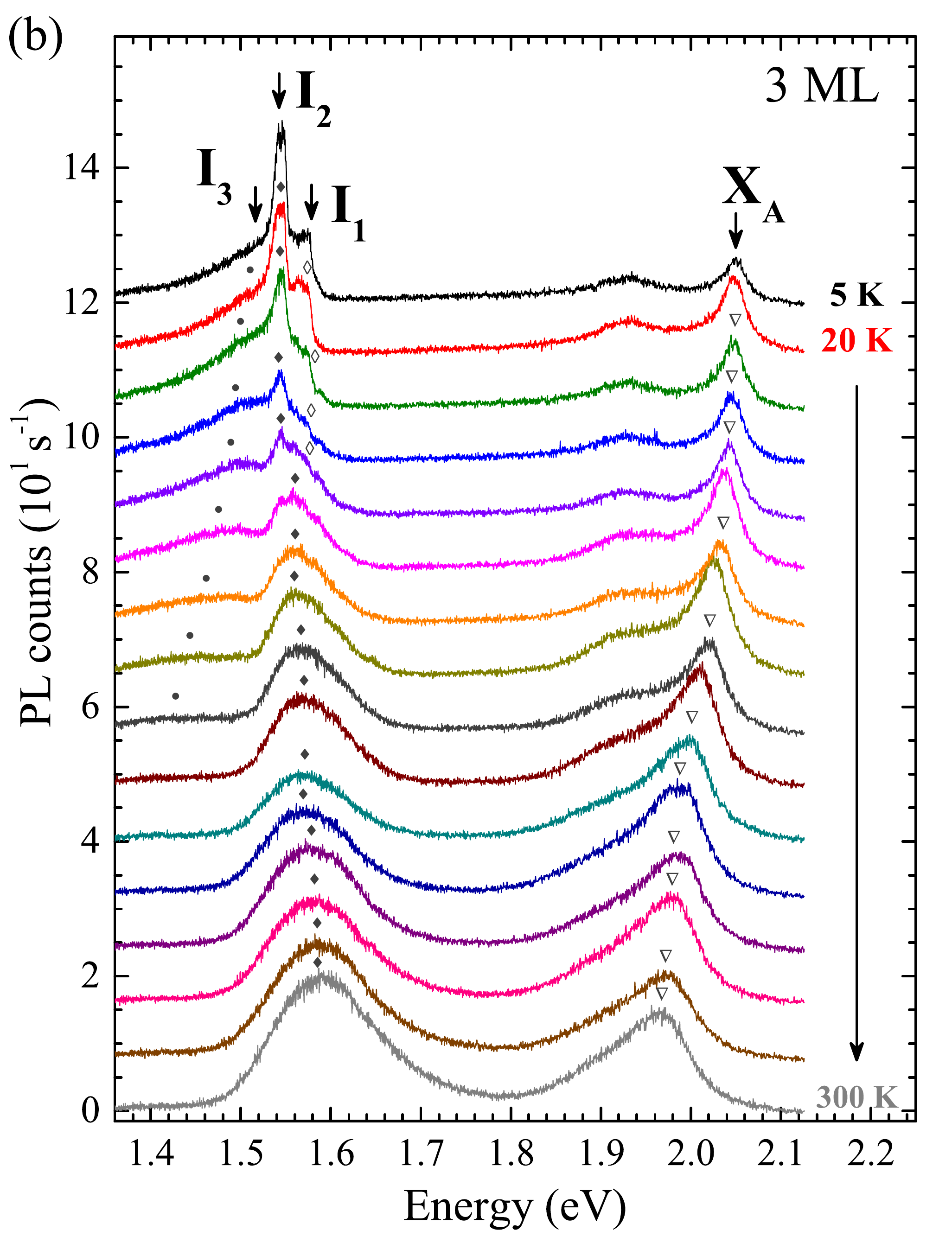}%
		\caption{Temperature evolution of photoluminescence spectra measured with the use of 2.41 eV laser light excitation on (a) 2 ML- and (b) 3 ML-thick flakes of WS$_2$. The spectra are vertically shifted for clarity.}
		\label{fig:PLvsT_2_3_ML}
	\end{figure*}
\end{center}

As discussed above the intense PL band of monolayer WS$_2$, related
to the direct optical transitions at the K$^\pm$ points of the BZ spans 
the energy range between 1.95~eV and 2.1~eV. The traces of
direct band-gap emission are still apparent in the spectra of bi-
and tri-layer WS$_2$, but the most pronounced PL band in all our
multilayers appears at significantly lower energies. This latter
emission band, logically assigned to indirect recombination
process, progressively shifts with $N$ towards lower energies. As
shown in Fig.~\ref{fig:PL_vs_Thick}, the indirect PL signal is
composed of two well defined peaks we will refer to as I$_1$ (the higher-energy one) and I$_2$ (the lower-energy one) and an additional broader band, I$_3$, which is quite pronounced in the spectrum of 2-ML-thick flake
but progressively weakens with $N$. The I$_1$ and I$_2$ features are assigned to the $\Lambda$-$\Gamma$
electron-hole recombination processes ($\Lambda$-$\Gamma$ indirect 
excitons) whereas we are less confident about the origin of the I$_3$
emission band. This is more discussed below in the context of the
results of the PL measurements performed as a function of temperature. 
The characteristic pattern of I$_1$ and I$_2$
emission peaks, separated in energy by 30~meV (see Fig.~\ref{fig:EnInd_vs_thick}), is
remarkably well reproduced in all our multilayers. This makes us
tempting to assume that these peaks are rather due to phonon-assisted and not
disorder-activated, $\Lambda$-$\Gamma$ recombination
processes. Apart from the case of low energy interlayer modes, the
spectrum of phonons is practically the same for all multilayers
whereas the effects of disorder are expected to be different in
different samples. We suppose that the I$_1$ and I$_2$ recombination
processes involve two different phonons characterized by energies that differ by
30~meV and the same k-vector corresponding to the $\Lambda$
point of the BZ (to match the wave vector difference of
the recombining $\Lambda$-electrons and $\Gamma$ holes). Pairs of such
phonons composed of one coming from the acoustic branch and the second
one from the optical branch, can be indeed found in the calculated
phonon dispersion relations of WS$_2$ layers.\cite{molina} Our rather qualitative
reasoning is, however, not sufficient to decide which particular
acoustic (TA or LA) and optical (of E- or A-symmetry) phonons are involved in
the I$_1$ and I$_2$ recombination processes. The problem remains to be
solved with the help of possible, future theoretical works.

Particularly interesting is the way the energies of the $\Lambda$-$\Gamma$ emission
peaks change with $N$ in our $N$-ML structures (see Fig.~\ref{fig:EnInd_vs_thick}). The indirect
$\Lambda$-$\Gamma$ emission peaks shift towards lower energies, from about 1.76~eV for 2 MLs 
to $\sim$1.4~eV for 8~MLs. The exact form of this 
dependence is ultimately determined by the changes in the band structure and in the
exciton binding energy in multilayer WS$_2$. A detailed examination of these
effects definitely calls for a thorough theoretical study. On the other hand
it is quite apparent that the derived experimental points can be
well reproduced with a rather simple formula which applies to the
energy evolution of the 2D confined electronic states in a
rectangular well with infinite barriers. The dashed curve in Fig.~\ref{fig:EnInd_vs_thick}
follows the formula:
\begin{equation}
E(L)=E_0 + \alpha{\slash}N^{2} = E_0 + \pi^2\hbar^2/2L^2\mu_{\perp},
\label{eq:qw}
\end{equation}
in which we have set $E_0$=1.39~eV and $\alpha$=1.71~eV or, conversely,
$\mu_{\perp}=0.58\hspace{0.5mm}m_{0}$, with $m_{0}$ being the free electron mass, when 
assuming $L=Na$ where $a$= 0.61615~nm\cite{schutte} is the thickness of
the WS$_2$ monolayer. Evoking the latter form of the above equation to
describe the observed $E$ versus $L$ dependence we are inclined to
consider that this dependence is mainly determined by the combined
effect of the energy shift of the 2D confined subbands in the
conduction and valence bands. $\mu_{\perp}$ should be then interpreted as
a reduced effective mass $\mu_{\perp}=(1/m_e+1/m_h)^{-1}$,  where $m_e$ ($m_h$) is the effective mass
which defines the parabolic dispersion relation along the c-axis
of bulk WS$_2$ for electrons at the $\Lambda$ point (holes at $\Gamma$ point ) of
the BZ. $E_0$=1.39~eV is the extrapolated energy of the
indirect PL transition in the bulk WS$_2$, indeed not far away
from the reported value of the indirect band gap in bulk WS$_2$ (Ref. \citenum{julien}).

The assignment of the I$_1$/I$_2$ emission band to the
$\Lambda$-$\Gamma$ indirect recombination processes is
confirmed by the characteristic evolution of these peaks
upon increasing the temperature. As illustrated in Fig.~\ref{fig:PLvsT_2_3_ML}
with the data measured for 2 ML and 3 ML flakes, an increase
of temperature results in the shift of the I$_1$/I$_2$ 
emission band towards higher energies. This is in 
contrast with a common behaviour characteristic of most
of semiconductors and observed also for 
the direct band gap (X$_\textrm{A}$) luminescence in our 
WS$_2$ flakes. Band edges in semiconductors shift with 
temperature mostly because of thermal expansion of
the crystal, which usually results in the shrinkage of 
the band gap. A conventional shrinkage of the direct band gap 
in S-TMD structures that occurs between the K-point electronic states, which
are well localized within the individual layers, results 
from the temperature expansion of those layers in lateral directions. Instead, the crystal expansion across 
the layers leads to a larger separation between the layers. 
This logically accounts for the blue shift of the indirect 
band gap which may eventually overcome the direct band gap 
in the limit of decoupled layers (set of separated monolayers).\cite{zhao2012,zhao2013} 
Another sign of the phonon-mediated indirect recombination
is a characteristic higher energy tailing of the I$_1$/I$_2$ band, 
which clearly develops upon increasing the temperature. At high 
temperatures, the electrons and holes occupy higher energy 
states with larger k-vectors, but any electron-hole pair 
can effectively recombine in the assistance of a phonon with 
an appropriately matched k-vector. Similar arguments apply 
to indirect excitons: there is \textit{a priori} no restriction
for the phonon-assisted recombination of indirect excitons 
with a non-zero center-of-mass K-vector. The population of such excitons increases
with temperature thus leading to the effective emission at higher energies.

As mentioned above, the origin of the broad I$_3$ band is less clear. 
Its overall energy position indicates that it is associated 
with an indirect transition but, on the other hand, it displays
a red shift with increasing the temperature, similar to the one characteristic
of the direct band-gap transitions (X$_\textrm{A}$) (see Fig.~\ref{fig:PLvsT_2_3_ML}). 
The I$_3$ emission band is effectively quenched with temperature
which may indicate its defect-mediated character. Following the 
previous studies (Ref. \citenum{zhao2013}), we speculate
that the I$_3$ emission might be due to indirect recombination
process but involving the defect states pinned to the CB edge at 
the K point and/or to the VB states at the $\Gamma$ point of the BZ.
This can very qualitatively account for the red shift of the I$_3$ band
(following that of the CB at the K point\cite{zhao2013}) as well
as its energy position below the I$_1$ and I$_2$ emission peaks
(large binding energies of defects at the K point).

\section{Absorption resonances in WS$_2$ multilayers }\label{sec:absorption}

\begin{center}
	\begin{figure}[t]
		\centering
		\includegraphics[width=0.81\linewidth]{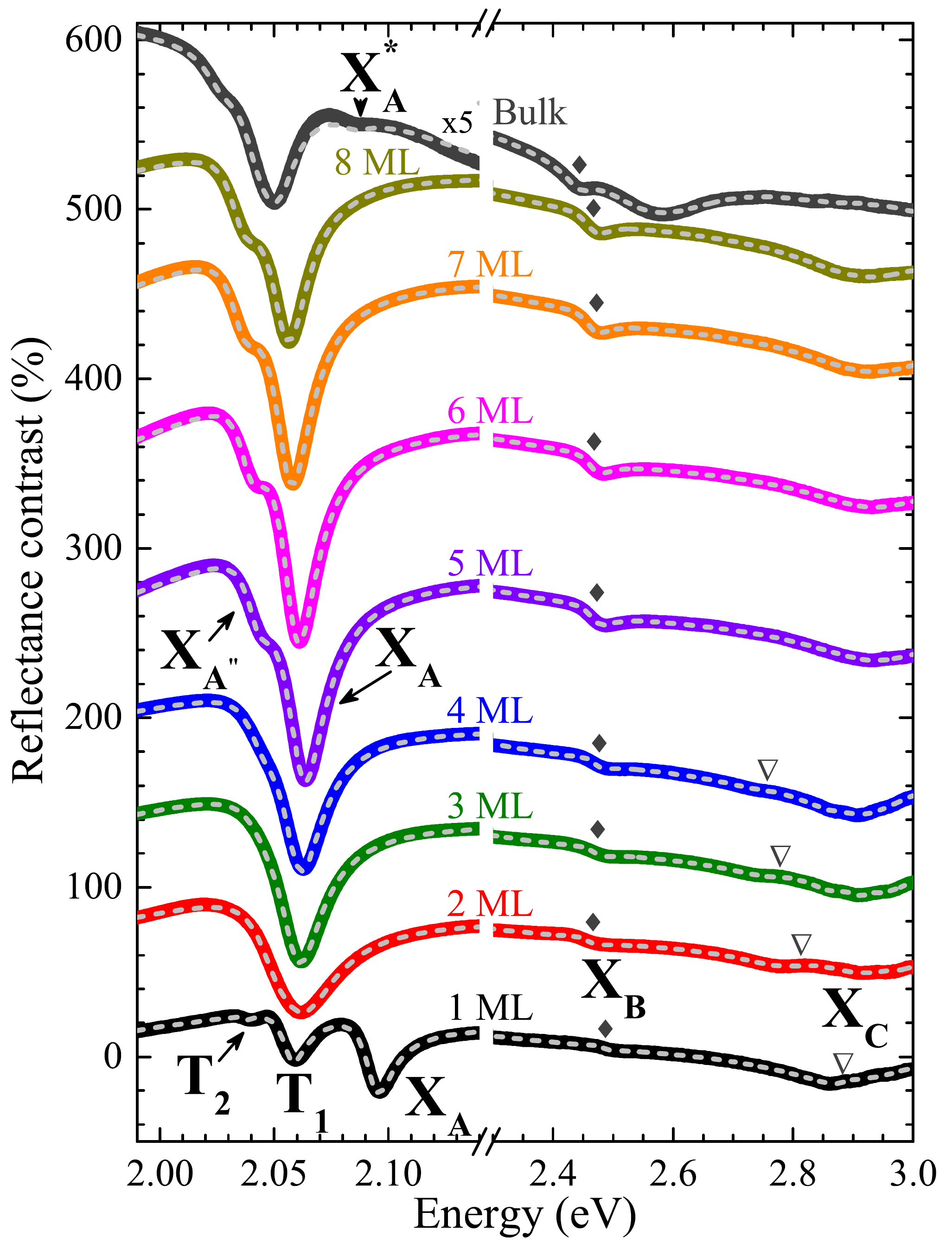}%
		\caption{Reflectance contrast spectra of WS$_2$ flakes shown in Fig.~\ref{fig:PL_vs_Thick} and of the bulk-like flake of 32~nm thickness measured at $T$=5~K. Drawn with dashed grey lines are corresponding model curves obtained with the aid of transfer matrix method combined with the Lorentz oscillator model. The spectra are vertically shifted for clarity.}
		\label{fig:RvsThick}
	\end{figure}
\end{center}

Low-temperature ($T$=5~K) RC spectra of our WS$_2$ flakes 
measured in a wide spectral range are shown in Fig.~\ref{fig:RvsThick}. 
The spectra display a pronounced group of resonances (A-exciton region) 
in the vicinity of $E$=2.05~eV, as well as somewhat weaker and broader
features at higher energies that arise from the B-exciton resonance (X$_\textrm{B}$)
at the K-point of the BZ and from another resonance (X$_\textrm{C}$) which is possibly 
related to a direct optical transition that takes place apart from 
the K points.\cite{zhao2012,zhu,li}

\begin{center}
	\begin{figure}[t]
		\centering
		\includegraphics[width=0.7\linewidth]{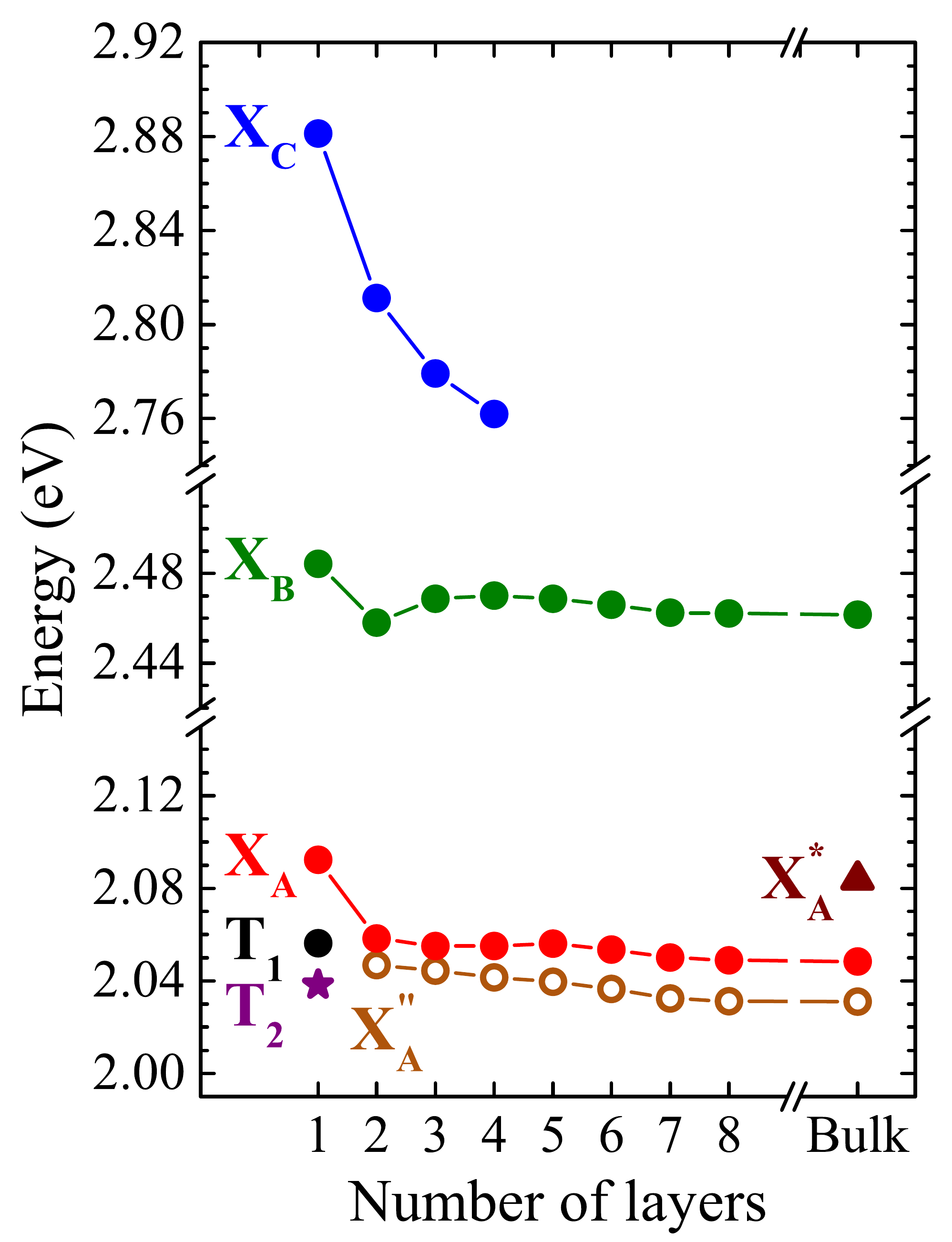}%
		\caption{Thickness dependence of the energy positions of excitonic resonances shown in Fig.~\ref{fig:RvsThick}. }
		\label{fig:EnR_vs_thick}
	\end{figure}
\end{center}

To analyze these spectra in more detail, a transfer matrix method combined with the Lorentz oscillator model 
have been employed to simulate the RC response of our flakes (see dashed grey
curves in Fig.~\ref{fig:RvsThick}). Bearing in mind a phenomenological approach, our intention was to simulate the data in a coherent way for all mutlilayers ($N=2$ to 8) and, at the same time, to keep the number of adjustable parameters, like the number of resonances in particular, at a minimal level. Importantly, the spectral response in the region of the A-exciton cannot be satisfactorily reproduced with a single resonance in most of our structures, $i.e.$ not only in the previously discussed case of the monolayer but also in $N$-MLs with $N \geq 3$ as well as in the bulk-like sample (see Fig.~\ref{fig:RvsThick} for a side band appearing on the low-energy slope of the main feature present around 2.05 eV, especially for $N \geq 5$). Two resonances, X$_{\textrm{A}}$ and X$_{\textrm{A"}}$, have to be actually taken into account in order to correctly simulate the group of A excitons in all our $N$-MLs with $N \geq 3$. The only ambiguous case is the bilayer, where the difference between the quality of fits obtained under the assumptions of single- and double-resonance character of the A-exciton group is purely quantitative (refer to Sec. III of the ESI for details). For the bulk-like flake of 32 nm thickness one more resonance denoted by X$_{\textrm{A*}}$ has to be included in the simulations to account for the first excited state of the A exciton that occurs in the spectrum around 2.09 eV.

The energy positions of the absorption resonances established 
in the course of modelling the RC response of our WS$_2$ multilayers 
are presented in Fig.~\ref{fig:EnR_vs_thick}. A prominent red shift,
as a function of number of layers, of the energy position of the X$_C$ resonance
indicates the pronounced changes in the electronic bands associated 
with this transition. Strikingly, however, the energy positions of the group 
of A-excitons and of the X$_\textrm{B}$ resonance show a rather weak dependence on $N$. As already known, the exciton binding energy in S-TMD multilayers exhibits a significant variation with the number of layers. 
It falls in the range of a few hundreds of meVs in the WS$_2$ monolayer,\cite{chernikov,zhu,ye} 
but amounts to only a few tens of meVs in bulk WS$_2$.\cite{chernikov} A weak dependence
on $N$ of the energy position of the A and B excitonic resonances implies therefore 
an approximate compensation of the reduction of the exciton binding energy 
by the shrinkage of the band gap. Both A- and B-excitons are associated 
with the K-point band-edge electronic states which are well localized 
in individual layers. Thus, a change in the dielectric screening, expected 
as a function of $N$, might be a dominant factor which leads to the renormalization
of both the single-particle band gap and the exciton binding energy.\cite{Ugeda2014} 
It seems, however, that these parameters are modified in a similar manner, 
making the actual energy positions of excitonic resonances virtually independent of $N$.

The multiple-resonance character of the A-exciton feature revealed in most of our
$N$-MLs is an intriguing observation. As extensively discussed in the
preceding section, it can be well understood in the case of the monolayer 
spectra which in addition to features originating from neutral excitons also display other resonances
due to charged excitons. The appearance of charged A- or B-excitons can be 
envisaged in multilayer WS$_2$ as well, though such three-particle complexes, not identified so far,
would necessarily be composed of an exciton (an electron-hole pair) at the K-point of the BZ
and a surplus quasi-particle (an electron or a hole) located at the minimum (maximum) 
of the CB (VB) at the $\Lambda$ ($\Gamma$) point of the BZ of 
multilayer WS$_2$. The bandwidth of A-excitons' response 
increases with $N$, what is clearly accounted for by the increase of the separation 
between the X$_\textrm{A}$ and X$_\textrm{A"}$ resonances deduced from 
the simulation of the RC spectra (see Fig.~\ref{fig:EnR_vs_thick}). This is 
in contrast with what one would expect for the binding energy of charged excitons, 
which should be a fraction of the binding energy of the neutral exciton, and
thus decrease with $N$. In consequence, it is rather unlikely that the charged A-excitons contribute to the RC spectra of our WS$_2$ multilayers.

\begin{center}
	\begin{figure}[t]
		\centering
		\includegraphics[width=0.95\linewidth]{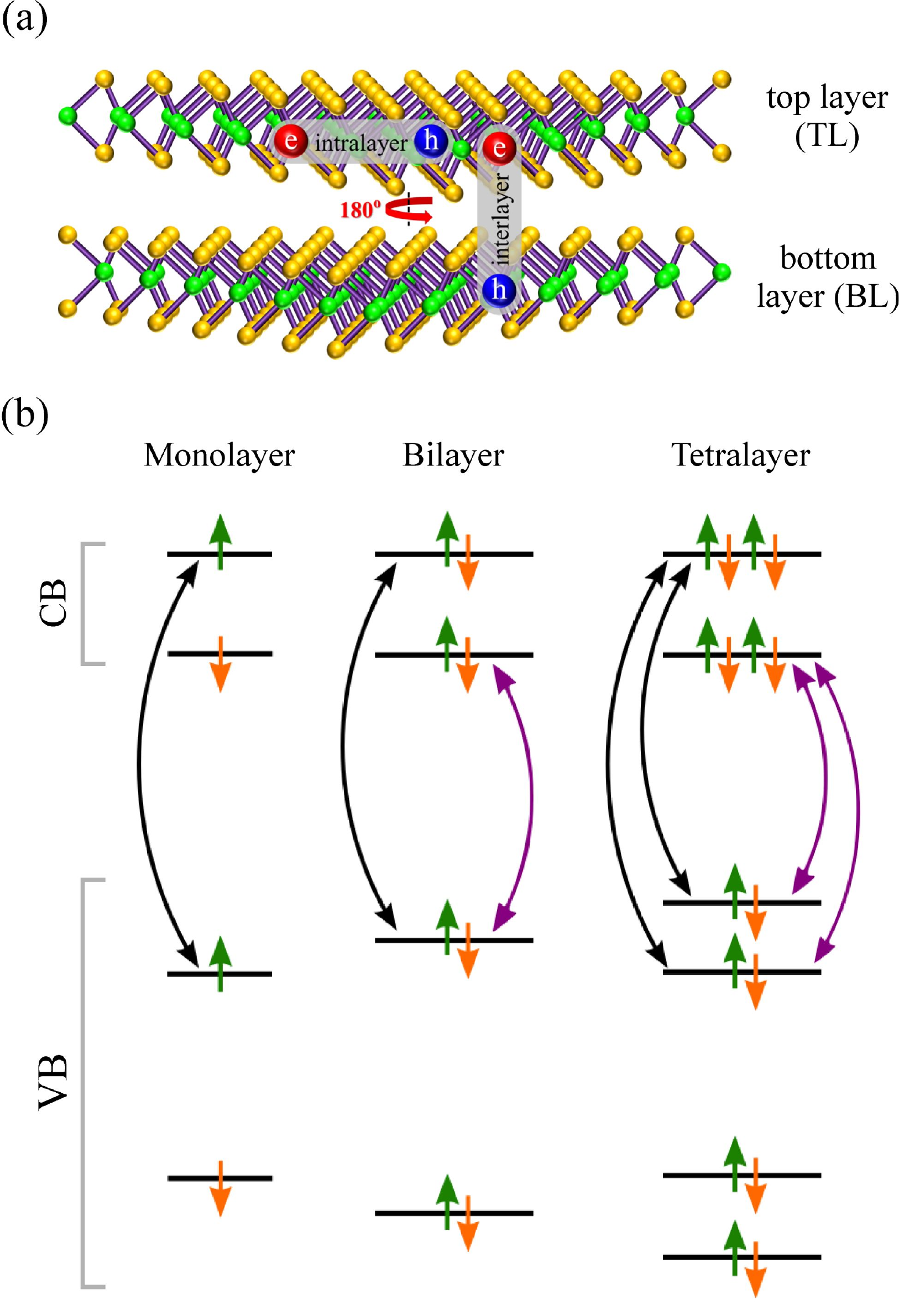}%
		\caption{(a) Schematic illustration of the layer configuration in a WS$_2$ bilayer, highlighting the intralayer and interlayer excitons. (b) Simplified diagram of the band structure of (left panel) monolayer WS$_2$, (middle panel) bilayer WS$_2$, and (right panel) tetralayer WS$_2$ at the K$^+$ point of the Brillouin zone. Small green (orange) arrows indicate the spin-up (spin-down) subbands. The black and purple arrows show the lowest-energy optical transitions due to intralayer and interlayer excitons, respectively.}
		\label{fig:theoryRC}
	\end{figure}
\end{center}

We rather presume that the multiple-, at least double-resonance character of the A-exciton band in our 2H-stacked $N$-MLs is an intrinsic effect which results from hybridization of the K-point electronic states and subsequent formation of multiple-component energy levels/subbands. Such an interpretation is discussed below on a qualitative level. Further details are given in respective sections of the ESI: Sec. II covering the theoretical background and Sec. IV presenting an approach to reproduce the experimental spectra.

In 2H-stacked films of WS$_2$ the arrangement of atoms in the direction across the layers is such that the metal atoms alternate with the chalcogen atoms in the subsequent layers. The geometry of the 2H stacking is schematically illustrated in Fig.~\ref{fig:theoryRC}(a) for a representative case of the bilayer. This geometry implies a particular arrangement of the atomic orbitals involved in the formation of electronic states in the k-space which leads to the alignment and resultant hybridization of the sequence of \mbox{K$^+$/K$^-$/K$^+$/K$^-$/$\ldots$} (\mbox{K$^-$/K$^+$/K$^-$/K$^+$/$\ldots$}) states originating from subsequent monolayers, which form the K$^+$ (K$^-$) band of the multilayer.  The symmetry of different, d$_0$ or d$_{\pm2}$ metal atom orbitals, implies that mixing of K-point electronic states is predominant for the valence band states (d$_{\pm2}$ orbitals) and encompasses the electronic states with the same spin orientation. The outcome of this mixing is a fine, multicomponent structure of each spin-orbit split subband of the VB at the K$^+$/K$^-$ points of the multilayer. Instead in the CB of $N$-ML , each spin-orbit split subband represents a set of $N$-degenerate (non-interacting) states (with different spins) from individual monolayers. The expected energy structure of K$^+$ electronic states for bi- and tetra-layer S-TMDs, together with that characteristic of the monolayer is schematically illustrated in Fig. \ref{fig:theoryRC}(b) (Apart of the inverted spin assignment, the K$^-$ states represent the identical energy diagram). The arrows shown in this figure account for the conceivable interband optical transitions (see ESI) associated with the upper spin-orbit split subband (A-exciton region). Two classes of optically active A-type transitions can be distinguished in our multilayers. One class of transitions comprises the resonances associated with the upper CB subband. These transitions (shown with black arrows in Fig. \ref{fig:theoryRC}(b)) display rather strong oscillator strength, comparable to that expected in monolayer, and involve the CB and VB states confined in the same layer (intralayer transitions/excitons). The transitions of the second class (purple arrows in Fig. \ref{fig:theoryRC}(b)) are associated with the bottom CB subband and are reminiscent of the spin forbidden transition in the monolayer. They involve the CB and VB states, each one largely confined in different (adjacent) layers (interlayer transitions/excitons) and display of about 10 times weaker oscillator strength than the intralayer transitions. Importantly, the multiplicity of possible A-type transitions as well as their spread in energy grows with the number of layers. This is in accordance with our phenomenological analysis of the RC data which shows the enlarged, with $N$, bandwidth of the A-exciton response (increased separation between X$_\textrm{A}$ and X$_\textrm{A*}$ resonances). It should be stressed out that our reasoning to account for the non-trivial structure of the A-resonances in WS$_2$ multilayers refers only to a single-particle picture (interband transitions) but neglects the otherwise important effects of Coulomb interactions. Those interactions may significantly modify our suggestions regarding the energy diagrams and oscillator of the A-type absorption resonances in WS$_2$ multilayers. Nonetheless, the approach presented in the ESI to calculate the RC spectra of multilayer WS$_2$ in the A-exciton region, which follows our simple theoretical model, fairly well reflects the measured spectra of $N$-ML WS$_2$ with $N$=2 to 8. 

\section{Conclusions}\label{sec:conclusions}

Concluding, we have presented the systematic studies of the optical response (absorption-like and photoluminescence spectra) of a series of WS$_2$ structures with different thickness: mono- and 2H-stacked multi-layers (up to 8 layers) as well as a 32~nm-thick, bulk-like sample. The fundamental absorption edge of the WS$_2$ monolayer has been found to be determined by the neutral exciton and, unusually, by the two distinct charged exciton resonances. The relative strength of these resonances has been investigated as a function of the electron concentration and temperature. A formation of two different charged excitons involving an access electron either from the bottom- or the upper- spin orbit-split subband of the WS$_2$ conduction band has been proposed. The recombination processes mediated by impurities (donor-acceptor and bound exciton recombination),  with their competing efficiencies, have been proposed to account for the observed nonlinearities, as a function of the excitation power, of the PL spectra of monolayer WS$_2$ measured at low temperatures, as well as for the characteristic evolution of the PL with temperature. The predominant, well-defined in-shape photoluminescence signal in WS$_2$ multilayers has been identified as due to phonon-mediated indirect recombination processes. The energy position of the indirect emission, when traced as a function of number of layers, has been found to follow a simple model of the confinement of electronic states in a rectangular quantum well with infinite barriers. The investigations of the absorption-like spectra in WS$_2$ multilayer demonstrate weak dependence of the energies of the fundamental A- and B-resonances upon $N$. This indicates an approximate compensation of the reduction of the exciton binding energy by the shrinkage of the bandgap and is likely a consequence of the dielectric screening effects in two-dimensional geometry. The fundamental direct bandgap absorption (A-exciton region) has been found to display a non-trivial, multiple resonance character. On the phenomenological level, the measured reflectance spectra of can be well reproduces assuming a double resonance structure of the A-exciton in any multilayer. Nonetheless, the discussed framework of the hybridization of the electronic states at the K-points of the Brillouin zone of the WS$_2$ multilayer indicate a more complex scheme of the A-exciton resonance in these systems. Extra information about the temperature activated shifts of the emission and absorption bands in our WS$_2$ structures are provided in the ESI.

%%%END OF MAIN TEXT%%%

\section*{Methods}\label{sec:methods}

Monolayer and few-layer flakes of WS$_2$ were obtained on a degenerately doped \mbox{Si/(320\hspace{0.5mm}nm)\hspace{0.5mm}SiO$_2$} substrate by polydimethylsiloxane-based exfoliation\cite{gomez} of bulk WS$_2$ crystals (2H phase) purchased from HQ Graphene. The flakes of interest were first identified by visual inspection under an optical microscope and then subjected to atomic force microscopy and Raman spectroscopy characterization in order to unambiguously determine their thicknesses. The back-gated WS$_2$ monolayer structure was fabricated by means of laser lithography performed on one of exfoliated flakes spin-coated with positive-tone photoresist. After exposure and development of the photoresist, the electrical contacts to the flake were defined by electron-beam evaporation of Ti/Au stacks, followed by a standard lift-off process, and annealing of the final device at 180$^{\circ}$C for 2 hours on a hot plate kept in air. During the measurements the WS$_2$ monolayer was grounded and the positive or negative electrical potential was applied to the silicon substrate via its freshly cleaved edge contacted by a drop of silver paint. The \textmu-PL measurements were carried out using 514.5 nm radiation from a continuous wave $Ar^+$ laser. The investigated sample was placed on a cold finger in a continuous flow cryostat mounted on x-y motorized positioners. The excitation light was focused by means of a 50x long-working distance objective with a 0.5 numerical aperture producing a spot of about 1~{\textmu}m diameter. The PL signal was collected via the same microscope objective, sent through a 0.5 m monochromator, and then detected by a liquid nitrogen cooled charge-coupled device camera. The excitation power focused on the sample was kept at 50 \textmu W during all measurements to avoid local heating. The \textmu-PLE measurements were carried out with the use of a tunable dye laser operating on rhodamine 6G. The PLE traces were obtained by recording variations of the intensity of light emitted at a particular energy while sweeping the excitation energy. In the case of the \textmu-RC study, the only difference in the experimental setup with respect to the one used for recording the \textmu-PL signal concerned the excitation source, which was replaced by a 100 W tungsten halogen lamp. The light from the lamp was coupled to a multimode fiber of a 50~\textmu m core diameter, and then collimated and focused on the sample to a spot of about 4~{\textmu}m diameter. The PL and RC measurements were performed at temperatures ranging from 5 K to 300 K.

If $R(E)$ and  $R_0(E)$ are the energy dependent reflectance spectra of the WS$_2$ flake and of the Si/SiO$_2$ substrate, respectively, then the percentage reflectance contrast spectrum is defined as follows:

\begin{equation}
RC(E)=\frac{R(E)-R_0(E)}{R(E)+R_0(E)}\times 100\%.
\label{eq:contrast}
\end{equation}

For the lineshape analysis of the RC spectra, we followed a method similar to that described in Ref. \citenum{arora}. We considered the excitonic contribution to the dielectric response function to be given by a modified Lorentz-oscillator-like model as:

\begin{equation}
\varepsilon(E)=(n_b(E)-ik_b(E))^2+\sum_p\frac{A_p}{E_p^2-E^2-i\gamma_pE},
\label{lorentz}
\end{equation}

where $E$ is the energy, and $n_b(E)+ik_b(E)$ represents the background complex refractive index of WS$_2$ in the absence of excitonic resonances, which in the simulations was assumed to be equal to that of the bulk material\cite{beal1976}. The index $p$ stands for the type of excitonic resonance characterized by a transition energy $E_p$, an amplitude $A_p$, and a phenomenological broadening parameter $\gamma_p$ (equal to full width at half maximum of the Lorentzian function). The refractive indices of Si and SiO$_2$ were obtained from Refs \citenum{si,sio2}. In order to ensure good correspondence between our simulations and the experimental conditions, we took into account the incidence of light on the sample surface at arbitrary angles enclosed within a solid angle defined by the numerical aperture of the objective used. We also included in the calculations a Gaussian distribution of power within the spot of exciting light, as well as an averaging over two possible polarizations (s and p) of the incident light. When simulating the RC spectra using a transfer matrix method, the thickness of WS$_2$ flakes was set either to a corresponding multiple of 0.61615 nm\cite{schutte} (the experimental thickness of a single layer) for few-layer flakes or to 32 nm for the bulk-like flake. Although treated as a fitting parameter, the thickness of SiO$_2$ layer stayed very close to 320 nm for all the cases considered. With the aid of this approach we were able to nicely reproduce the shape of the RC spectra not only in the ranges dominated by excitonic resonances but also apart from them.

\section*{Acknowledgements}

The work has been supported by the European Research Council (MOMB project no. 320590), the EC Graphene Flagship project (no. 604391), the National Science Center (grant no. DEC-2013/10/M/ST3/00791), the Nanofab facility of the Institut N\'eel, CNRS UGA, and the ATOMOPTO project (TEAM programme of the Foundation for Polish Science co-financed by the EU within the ERDFund).

M. R. M. and K. N. contributed equally to this work.

\bibliographystyle{apsrev4-1}
\bibliography{biblio}

\newpage
\onecolumngrid
\setcounter{figure}{0}
\setcounter{section}{0}
\renewcommand{\thefigure}{S\arabic{figure}}
	\begin{center}
	%%%%%%%%% ABSTRACT TITLE
	{\large{{\bf  \textsc{Electronic Supporting Information}} \\ Optical response of monolayer, few-layer and bulk tungsten disulfide}}
	%%%%%%%%% ABSTRACT AUTHORS
	\vskip0.5\baselineskip{Maciej R. Molas,{$^{1,\ast}$} Karol Nogajewski,{$^{1,\dag}$}, Artur O. Slobodeniuk,{$^{1}$} \linebreak[4] Johannes Binder,{$^{1,2}$} Miroslav Bartos,{$^{1}$} and Marek Potemski,{$^{1,2,\ddag}$}}
	%%%%%%%%% AFFILIATION
	\vskip0.5\baselineskip{\em$^{1}$ Laboratoire National des Champs Magn\'etiques Intenses, CNRS-UGA-UPS-INSA-EMFL, 25, avenue des Martyrs, 38042 Grenoble, France \\$^{2}$ Faculty of Physics, University of Warsaw, ul. Pasteura 5, 02-093 Warszawa, Poland }
	\end{center}

\footnotetext{\textit{E-mail: maciej.molas@gmail.com, karol.nogajewski@lncmi.cnrs.fr,marek.potemski@lncmi.cnrs.fr}}
\footnotetext{\ddag~These authors contributed equally to this work}

\section{Influence of the monolayer PL on the emission of light from thicker WS$_2$ films}

It is worth to point out an important property of selected samples prepared by means of mechanical exfoliation, which may significantly affect their PL response. It particularly concerns the large-area flakes, which at the fabrication stage are subject to strongly non-uniform field of tearing force. Acting on thicker flakes, such force can locally split them into subunits composed of smaller number of layers, which nonetheless do not stop to interact with one another. For example, within a 3 ML-thick flake a pocket containing a 1 ML patch that is partially detached from its 2 ML-thick companion can be created. If optically probed, it may then give rise to the appearance of mixed PL signal comprising contributions of all three kinds, $i.e.$ the 1 ML-, 2 ML-, and 3 ML-like, of course distorted by interactions between the two subunits. When performing preliminary measurements we experienced such a situation for some of our flakes composed of 3~MLs and 4 MLs, whose PL spectra were dominated by a set of prominent features resembling those observed for the monolayer. In order to avoid influencing our results by this artifact, we then carefully examined all flakes under consideration before qualifying them for further optical study.   

\begin{figure}[h]
\centering
  \includegraphics[width=0.45\linewidth]{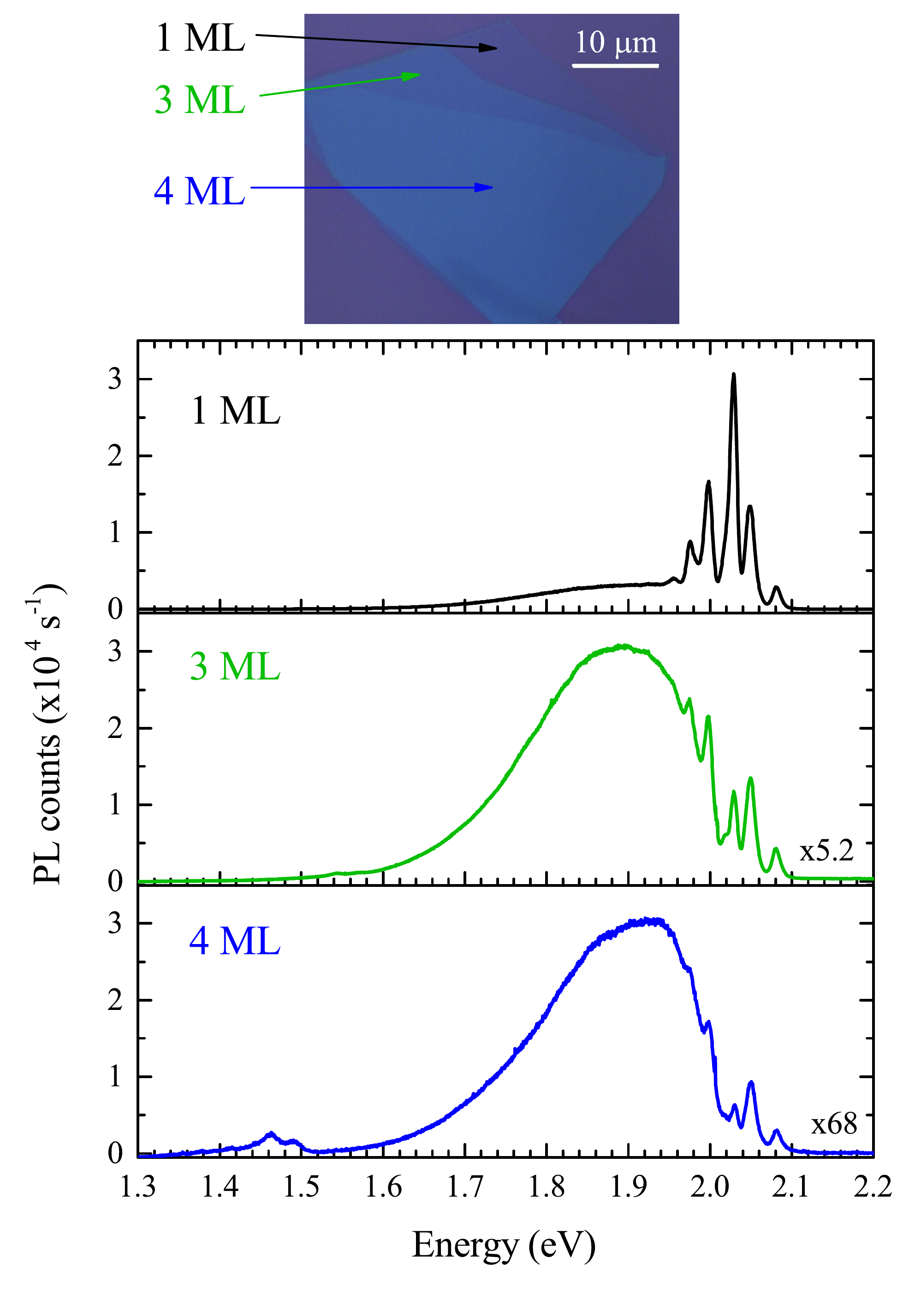}
  \caption{PL spectra of the 1 ML-, 3 ML-, and 4 ML-thick flakes of WS$_2$, measured at $T$=~5~K and the  excitation energy equal to 2.41~eV. The insets show the optical images of the studied flake.}
  \label{fig:PL_vs_Thick_SI}
\end{figure}

\newpage

\section{The band structure of S-TMD multilayers at the $\mathbf{K}$ points of the Brillouin zone}\label{sec:bndstrctr}

All the important features of multilayer S-TMDs exist also in a bilayer.
Therefore we examine the latter system in details (following Ref. \citenum{gong2013})
and then generalise our consideration to the multilayer
case. Finally we describe the properties of a bulk S-TMD.

\subsection{Bilayer S-TMD.}
We consider first the bilayer S-TMD as a system of two decoupled monolayers
with 2H stacking order. In this case, the bilayer Bloch states at the
$\mathrm{K}^\pm$ point can be constructed from $\mathrm{K}^\pm$ states of the
lower and  $\mathrm{K}^\mp$ states of the upper layers respectively (see Fig. 12b).

Let us focus on the $\mathrm{K}^+$ point of such system for clarity. Then, the
conduction and valence band states of the lower layer can be written as $\{|d^{(1)}_0\rangle\otimes|\uparrow\rangle,|d^{(1)}_0\rangle\otimes|\downarrow\rangle\}$ and
$\{|d^{(1)}_{+2}\rangle\otimes|\uparrow\rangle,|d^{(1)}_{+2}\rangle\otimes|\downarrow\rangle\}$.
The superscript ``$(1)$'' denotes the ``first'' (lower) layer and $s=\uparrow,\downarrow$
marks the spin-up and spin-down states respectively. The symbols $d_{m=0,2}$ show from which
transition metal atomic orbitals the quantum states are predominantly made. The conduction and
valence band states of the upper layer are $\{|d^{(2)}_0\rangle\otimes|\uparrow\rangle,|d^{(2)}_0\rangle\otimes|\downarrow\rangle\}$, $\{|d^{(2)}_{-2}\rangle\otimes|\uparrow\rangle,|d^{(2)}_{-2}\rangle\otimes|\downarrow\rangle\}$.
The superscript ``$(2)$'' denotes the ``second'' (upper) layer. The coordinate representation of the
$|d^{(1)}_{+m}\rangle$ and $|d^{(2)}_{-m}\rangle$ states can be found in Ref. \citenum{jones2014}.

\begin{figure}
	\hspace{-1.5cm}
	\includegraphics[width=9cm]{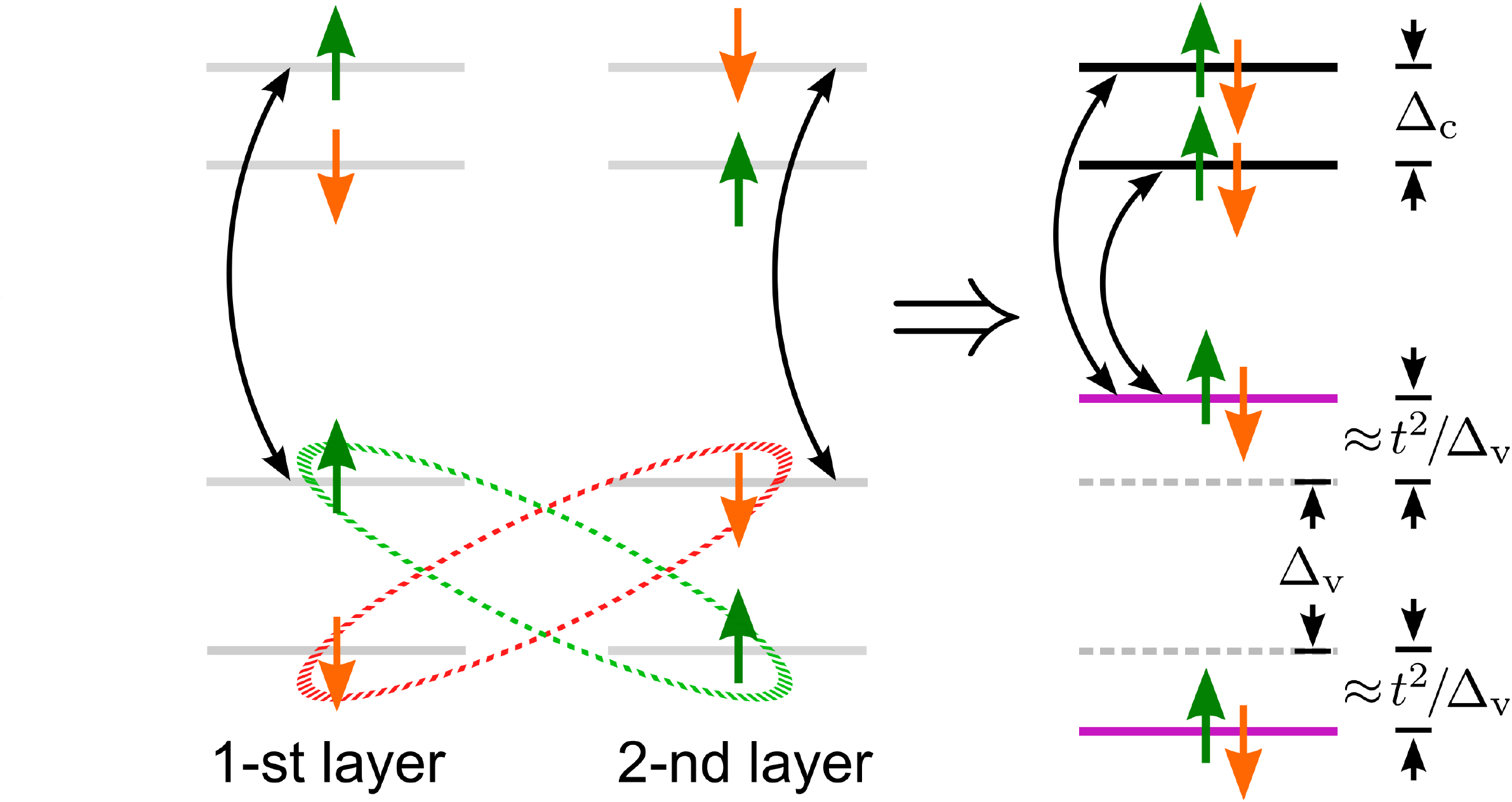}
	\caption{\label{fig:1}
		(Left panel) Diagram of the bands position of two 2H-stacked S-TMD monolayers
		(at the $\mathrm{K}^+$ point of the 1-st layer) in the absence of interlayer interaction.
		(Right part) Schematic drawing of the band positions in the bilayer, as a result
		of interaction between monolayers (presented by green and red dashed ellipses on the left panel).
		The green (orange) arrows indicate the spin-up (spin-down) subbands.
		Solid black arrow lines indicate the possible lowest energy optical transitions in both systems.}
\end{figure}

In the absence of interlayer hopping the valence and conduction bands at the $\mathrm{K}^+$
point are doubly degenerated, and each pair $\{|d^{(1)}_{+m}\rangle\otimes|s\rangle, |d^{(2)}_{-m}\rangle\otimes|-s\rangle\}$ for fixed $s$ and $m$ is characterized by the
same energy (see Fig.~\ref{fig:1}).
$|-s\rangle$ represents the spin state with the opposite to $s$ direction.
In fact, the degeneracy is lifted by interlayer forces, which mix the basis states.
The symmetry analysis~\cite{liu2015} shows that only the valence
band states with the same spin interact with each other.
The conduction bands remain doubly degenerated.

Let us study the band structure of bilayer S-TMD quantitatively.
We focus on the states at the $\mathrm{K}^+$ point for tungsten-based
compounds for brevity (the consideration of the opposite valley and/or
molybdenum-based structures can be done by analogy).

The conduction band of bilayer contains two doubly degenerated subbands
with the same splitting $\Delta_\mathrm{c}$ as in monolayer.
The sets of high- and low-energy states are $\{|d^{(1)}_0\rangle\otimes|\uparrow\rangle,|d^{(2)}_0\rangle\otimes|\downarrow\rangle\}$,
$\{|d^{(1)}_0\rangle\otimes|\downarrow\rangle,|d^{(2)}_0\rangle\otimes|\uparrow\rangle\}$.
The valence band states and their energies can be found from the
Hamiltonian~\cite{gong2013}
\begin{equation}
H_\mathrm{v}(\mathrm{K^+})=\left[
\begin{array}{cccc}
\Delta_\mathrm{v}/2 & t & 0 & 0\\
t & -\Delta_\mathrm{v}/2 & 0 & 0 \\
0 & 0 & \Delta_\mathrm{v}/2 & t \\
0 & 0 & t & -\Delta_\mathrm{v}/2 \\
\end{array}
\right],
\end{equation}
written in the basis $\{|d^{(1)}_2\rangle\otimes|\!\!\uparrow\,\rangle,|d^{(2)}_{-2}\rangle\otimes|\!\!\uparrow\,\rangle,
|d^{(2)}_{-2}\rangle\otimes|\downarrow\,\rangle, |d^{(1)}_2\rangle\otimes|\!\downarrow\,\rangle \}$.
The diagonal matrix elements $\pm\Delta_\mathrm{v}/2$ are the energies of corresponding states in the
absence of interlayer hopping, which is defined by parameter $t$.
According to block-diagonal form of Hamiltonian only the states with
the same spin are mixed. Both blocks are equal and, therefore has the same
spectrum. It results in two new doubly spin degenerated bands with energies
$E_\pm=\pm\sqrt{\Delta_\mathrm{v}^2/4+t^2}\approx \pm (\Delta_\mathrm{v}/2 + t^2/\Delta_\mathrm{v})$, see Fig.\ref{fig:1}.

The highest energy $E_+$ valence bands states of bilayer are
\begin{eqnarray}
|\Uparrow\rangle_+&=&\{\cos\theta |d^{(1)}_2\rangle+\sin\theta|d^{(2)}_{-2}\rangle\}\otimes|\uparrow\rangle,\\
|\Downarrow\rangle_+&=&\{\sin\theta|d^{(1)}_2\rangle+\cos\theta|d^{(2)}_{-2}\rangle\}\otimes|\downarrow\rangle,
\end{eqnarray}
where $\cos(2\theta)=\Delta_\mathrm{v}/\sqrt{\Delta_\mathrm{v}^2+4t^2}$.
The low-energy $E_-$ states can be calculated from $E_+$ ones by replacing
$\cos\theta\rightarrow -\sin\theta, \sin\theta \rightarrow \cos\theta$.

The obtained structure of the conduction and valence bands helps us to find the properties of lowest
energy optical transitions in bilayer S-TMD. There are four possible transitions of such type.
Two of them have the energy $E_\mathrm{X_A}-t^2/\Delta_\mathrm{v}$ and intensities of lines
\begin{equation}
I_1^\pm=I_0\cos^2\theta=I_0\left(\frac{1}{2}+\frac{\Delta_\mathrm{v}}{2\sqrt{\Delta_\mathrm{v}^2+4t^2}}\right)
\end{equation}
with $\sigma_\pm$ polarization of light.
$E_\mathrm{X_A}$ and $I_0$ are the energy and intensity of A exciton transitions in monolayer (solid black arrow lines,
which couple valence and conduction bands in left panel of Fig.~\ref{fig:1}).
The second pair of transitions has the energy $E_\mathrm{X_A}-\Delta_\mathrm{c}-t^2/\Delta_\mathrm{v}$ and intensity of lines
\begin{equation}
I_2^\pm=I_0\sin^2\theta=I_0\left(\frac{1}{2}-\frac{\Delta_\mathrm{v}}{2\sqrt{\Delta_\mathrm{v}^2+4t^2}}\right)
\end{equation}
with $\sigma_\pm$ polarization of light.
As one can see, the intensities of the left and right-polarized light in the case of
bilayer are the same. Therefore, this material doesn't have the valley dependent optical
dichroism as in monolayer S-TMD. This is due to the presence of the inversion symmetry in bilayer.

We estimate the energies and intensities of such transitions for the case of WS$_2$.
For WS$_2$ $t=0.0545$ eV, $\Delta_\mathrm{v}=0.421$ eV and $\cos\theta=0.992$, $\sin\theta=0.126$,
see Ref. \citenum{gong2013}. Hence the exciton energy shift, due to mixing in valence band,
is $t^2/\Delta_\mathrm{v}\approx7$ meV. The relative intensity $I^\pm_1/I^\pm_2\approx 60$. The same
optical properties the bilayer has in the vicinity of $\mathrm{K}^-$ point.

\subsection{Multilayer S-TMD.} 

A multilayer S-TMD crystal is a set of $\mathrm{N}$ monolayers,
arranged in a pile with 2H stacking order. As in the bilayer case we enumerate
all the S-TMD sheets from $1$ to $\mathrm{N}$, starting from the lowest (first) layer.
Then, according to the 2H stacking all even layers are $180^\circ$ rotated of the odd ones.
Therefore the basis valence and conduction states at the $\mathrm{K}^+$ point are  $\{|d^{(2m+1)}_{+2}\rangle\otimes|\uparrow\rangle,|d^{(2m+1)}_{+2}\rangle\otimes|\downarrow\rangle\}$, $\{|d^{(2m+1)}_0\rangle\otimes|\uparrow\rangle,|d^{(2m+1)}_0\rangle\otimes|\downarrow\rangle\}$ for odd layers,
and are $\{|d^{(2m)}_{-2}\rangle\otimes|\uparrow\rangle,|d^{(2m)}_{-2}\rangle\otimes|\downarrow\rangle\}$, $\{|d^{(2m)}_0\rangle\otimes|\uparrow\rangle,|d^{(2m)}_0\rangle\otimes|\downarrow\rangle\}$ for even ones,
where $m=1,2,\ldots [\mathrm{N}/2]$.

The interaction between layers is of the same type as in bilayer.
Namely, the conduction bands do not interact, and as a result form spin-up and spin-down
$\mathrm{N}$ degenerated conduction band states.
The valence band states are mixed due to interlayer hopping.
The corresponding valence band Hamiltonian is $2\mathrm{N}\times2\mathrm{N}$ block-diagonal matrix.
Each $\mathrm{N}\times\mathrm{N}$ block acts on spin-up and spin-down states separately.
The eigenvalues of the matrix defines the positions of new
valence bands.

Note that $\mathrm{N}=2\mathrm{M}$ (even) and $\mathrm{N}=2\mathrm{M}+1$ (odd) multilayers have different properties from spectroscopic point of view.~\cite{zeng2013}
For $\mathrm{N}=2\mathrm{M}$ ($i.e.$ 2~MLs, 4~MLs, 6~MLs,...), system possesses the inversion symmetry. Therefore, valley dependent physical properties of each
monolayer, such as optical selection rules, are averaging to zero in such systems. It reflects also
in the fact that for even number of layers all the new valence bands are, as minimum, doubly spin degenerated.

For $\mathrm{N}=2\mathrm{M}+1$ ($i.e.$ 3~MLs, 5~MLs, 7~MLs,...), there is one non-compensated additional layer in the crystal. It results in valley dependent optical dichroism in the system. However, it is suppressed by the large number of layers. The spectrum of such system contains the set of $\mathrm{M}$ doubly spin degenerated levels (very close by the structure to $2\mathrm{M}$ multilayer) and two additional levels with energies $\pm\Delta_\mathrm{v}/2$.

\begin{figure}
	\includegraphics[width=8cm]{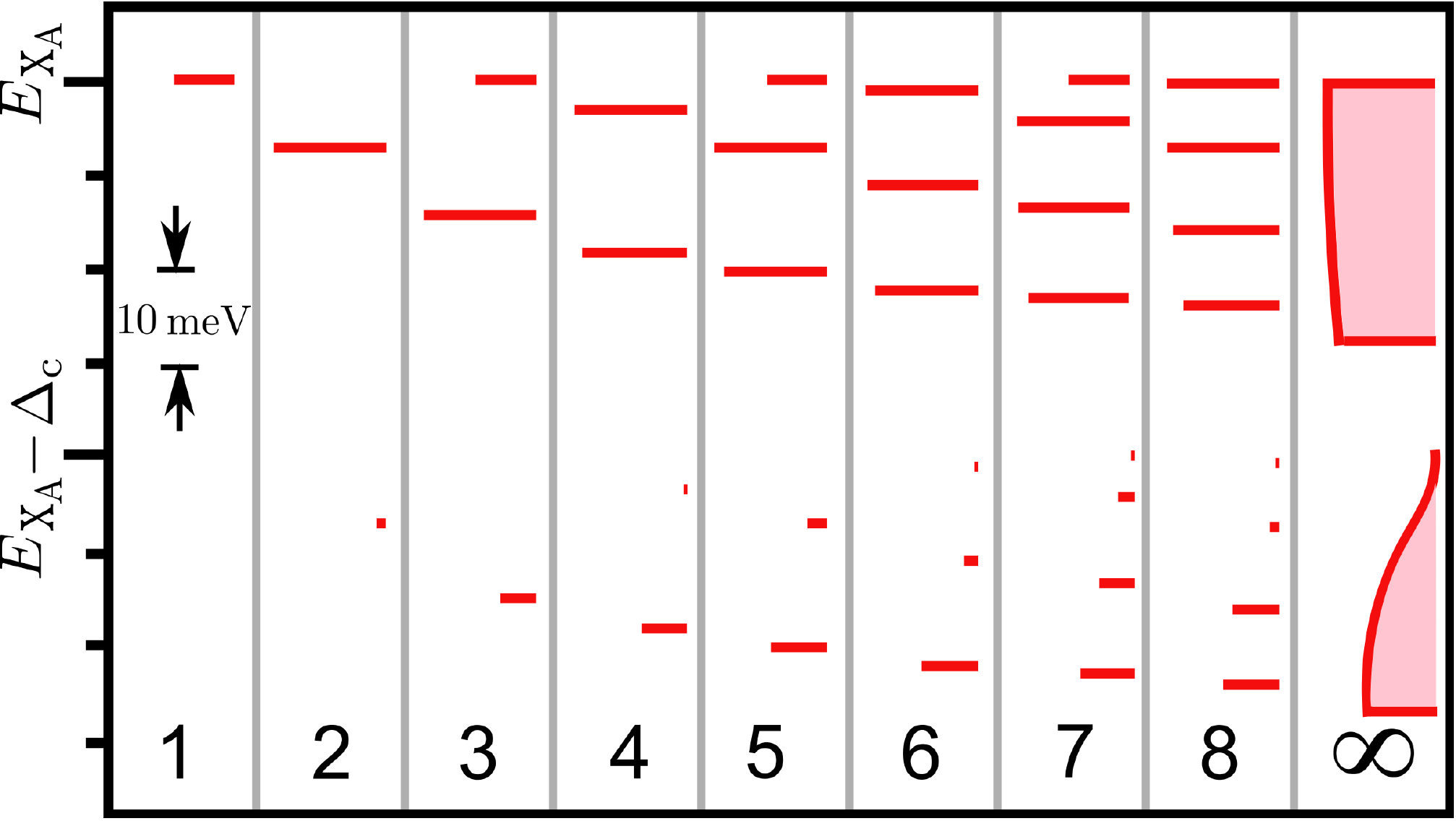}
	\caption{\label{fig:2}
		The position and relative intensities of lowest energy transitions in multilayer WS$_2$,
		as a function of number of layers. The length of each line from the top
		part of the picture is proportional to the intensity of corresponding transition. The
		intensity of transitions, which correspond to the bottom set of lines, is 10 times smaller
		than presented on the picture.}
\end{figure}

The optical intensities of excitonic transitions in multilayers can be found by solving the eigenvalue problem for $\mathrm{N}\times \mathrm{N}$ matrices, mentioned above. They are presented in Fig.~\ref{fig:2}.

The difference between odd and even multilayers disappears in the limit $\mathrm{N}\rightarrow\infty$.
Therefore, it is convenient to consider the bulk S-TMD as the even layered system $\mathrm{N}=2\mathrm{M}$.
Corresponding calculations are done in the next subsection.

\subsection{The band structure of bulk S-TMD.}

For a bulk S-TMD, according to previous consideration, there is no interlayer hopping for conduction states in
vicinity of $\mathrm{K}^\pm$ points, that reflects in spin-up and spin-down $\mathrm{N}$ degenerated
conduction bands. The valence band states interact with each other and form new bands.
In the bulk case $\mathrm{M}\rightarrow\infty$ all these new states belong to one wide band, which has
the meaning of the band in out-of-plane $(z)$ direction. Below we obtain this conclusion quantitatively.

The bulk S-TMD with $\mathrm{N}=2\mathrm{M}$ layers can be presented as a pile of $\mathrm{M}$ bilayers.
The length of periodicity
in such system coincides with the lattice constant $c$ of the bulk in out-of-plane $(z)$ direction.
Therefore we slice the bulk
on the elementary blocks with two layers inside, and then study the hopping between these blocks.
In further, we focus on the states at the $\mathrm{K}^+$ point in details (the case of $\mathrm{K}^-$
is described in the analogous way).

\begin{figure}
	\includegraphics[width=8cm]{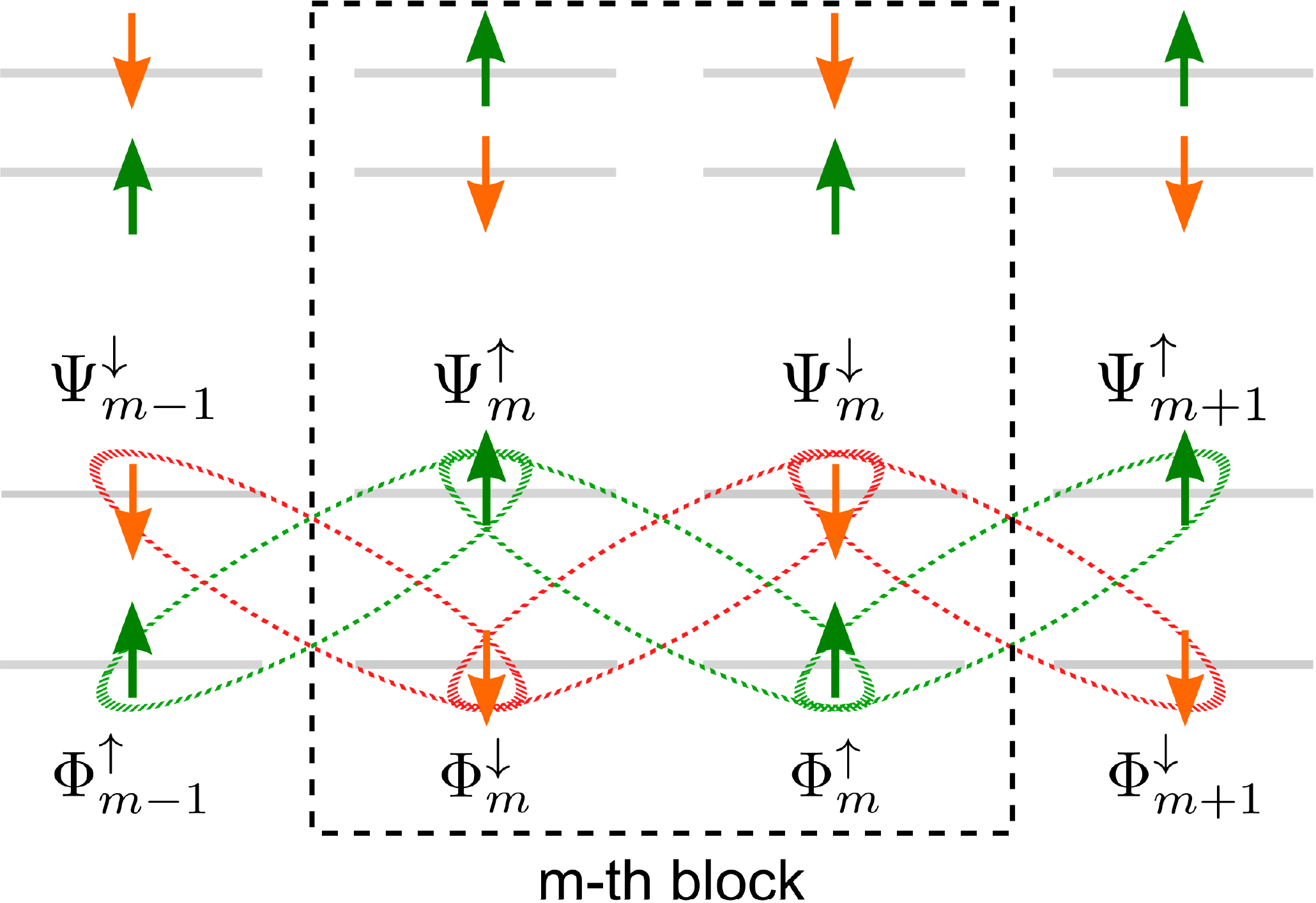}
	\caption{\label{fig:3}
		Schematic representation of the interaction
		between the states of $m$-th block and its neighbours
		at the $\mathrm{K}^+$ point of multilayer S-TMD. The interactions are denoted by green and red dashed ellipses.
		The green (orange) arrows indicate the spin-up (spin-down) subbands in the conduction and valence bands.}
\end{figure}

Let us consider the $m$-th bloc of sliced system, mentioned above (see Fig.\ref{fig:3}). Its valence bands basis
functions are $\{|d^{(m)}_2\rangle\otimes|\uparrow\rangle,|d^{(m)}_2\rangle \otimes|\downarrow\rangle,|d^{(m)}_{-2}\rangle\otimes|\uparrow\rangle,
|d^{(m)}_{-2}\rangle\otimes|\downarrow\rangle\}$.
The each state in the system we present as
\begin{eqnarray}
|\Uparrow\rangle=\frac{1}{\sqrt{\mathrm{M}}}\sum_{m=1}^\mathrm{M} \{\Psi_m^\uparrow|d^{(m)}_2\rangle +\Phi_m^\uparrow|d^{(m)}_{-2}\rangle\}\otimes|\uparrow\rangle, \\
|\Downarrow\rangle=\frac{1}{\sqrt{\mathrm{M}}}\sum_{m=1}^\mathrm{M} \{\Psi_m^\downarrow|d^{(m)}_{-2}\rangle +\Phi_m^\downarrow|d^{(m)}_2\rangle\}\otimes|\downarrow\rangle.
\end{eqnarray}
The variables $\Psi^s_m$ and $\Phi_m^s$ correspond to high and low-energy
excitations of valence band. The eigenvalue problem in this representation takes
the form
\begin{eqnarray}
E\Psi_m^\uparrow &=& \frac{\Delta_\mathrm{v}}{2}\Psi_m^\uparrow + t\Phi_m^\uparrow + t\Phi_{m-1}^\uparrow, \\
E\Psi_m^\downarrow &=&\frac{\Delta_\mathrm{v}}{2}\Psi_m^\downarrow + t\Phi_m^\downarrow + t\Phi_{m+1}^\downarrow, \\
E\Phi_m^\downarrow &=& -\frac{\Delta_\mathrm{v}}{2}\Phi_m^\downarrow + t\Psi_m^\downarrow + t\Psi_{m-1}^\downarrow, \\
E\Phi_m^\uparrow &=& -\frac{\Delta_\mathrm{v}}{2}\Phi_m^\uparrow + t\Psi_m^\uparrow + t\Psi_{m+1}^\uparrow.
\end{eqnarray}
The schematic representation of these relations is shown in Fig.~\ref{fig:3}

\begin{figure}[!b]
	\includegraphics[width=8cm]{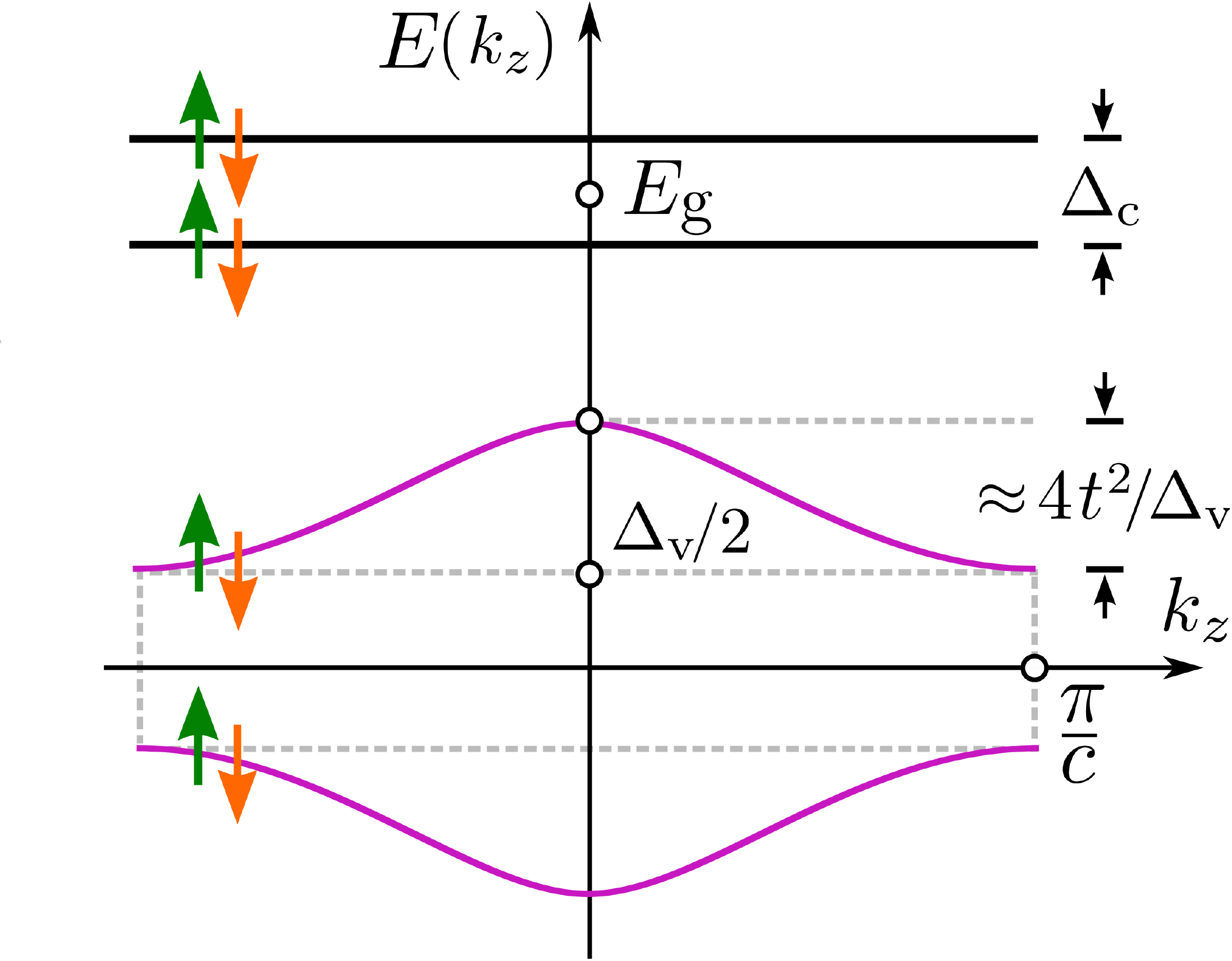}
	\caption{\label{fig:4}
		The valence (purple lines) and conduction (black lines) energy bands as a function of
		out-of-plane wave-vector $k_z$ at the $\mathrm{K}^+$ point. $E_g$ represents the position of band
		gap in monolayer S-TMD. The green (orange) arrows indicate the spin-up (spin-down) subbands in the
		conduction and valence bands.}
\end{figure}

Introducing  the substitution
\begin{equation}
\left[
\begin{array}{c}
\Psi_m^s \\
\Phi_m^s \\
\end{array}
\right]=e^{ikcm}\left[
\begin{array}{c}
\Psi_k^s \\
\Phi_k^s \\
\end{array}
\right]
\end{equation}
and periodic boundary conditions $k \in k_n=2\pi n/c\mathrm{M}, n\in(-[\mathrm{M}/2],[\mathrm{M}/2]]$, 
one present the relations between new states in the form
\begin{equation}
E(k)\left[
\begin{array}{c}
\Psi_k^\uparrow \\
\Phi_k^\uparrow \\
\end{array}
\right]
=\left[\begin{array}{cccc}
\Delta_\mathrm{v}/2 & 2te^{-i\frac{kc}{2}}\cos(\frac{kc}{2}) \\
2te^{i\frac{kc}{2}}\cos(\frac{kc}{2})& -\Delta_\mathrm{v}/2  \\
\end{array}\right]
\left[
\begin{array}{c}
\Psi_k^\uparrow \\
\Phi_k^\uparrow \\
\end{array}
\right]
\end{equation}
\begin{equation}
E(k)\left[
\begin{array}{c}
\Psi_k^\downarrow \\
\Phi_k^\downarrow \\
\end{array}
\right]
=\left[\begin{array}{cccc}
\Delta_\mathrm{v}/2 & 2te^{i\frac{kc}{2}}\cos(\frac{kc}{2}) \\
2te^{-i\frac{kc}{2}}\cos(\frac{kc}{2})& -\Delta_\mathrm{v}/2  \\
\end{array}\right]
\left[
\begin{array}{c}
\Psi_k^\downarrow \\
\Phi_k^\downarrow \\
\end{array}
\right]
\end{equation}
The spin-up and spin-down states are decoupled in this case, but they have the
same energy spectrum $E(k)=\pm\sqrt{\Delta_\mathrm{v}^2/4 + 4t^2 \cos^2(kc/2)}$.
This fact reflects that all new valence bands of the bulk are doubly spin-degenerated.
The corresponding band structure, in the limit $\mathrm{M}\rightarrow \infty$ ($k_n\rightarrow k_z$) is presented in Fig.~\ref{fig:4}.

Let us consider the higher energy bands $E(k_z)\geq \Delta_\mathrm{v}/2$ more precisely.
The corresponding eigenstates are
\begin{equation}
\left[
\begin{array}{c}
\Psi_{k_z}^\uparrow \\
\Phi_{k_z}^\uparrow \\
\end{array}
\right]=\left[
\begin{array}{c}
\cos\theta_{k_z} \\
e^{i\frac{{k_z}c}{2}}\sin\theta_{k_z} \\
\end{array}
\right], \quad
\left[
\begin{array}{c}
\Psi_{k_z}^\downarrow \\
\Phi_{k_z}^\downarrow \\
\end{array}
\right]=\left[
\begin{array}{c}
\cos\theta_{k_z} \\
e^{-i\frac{{k_z}c}{2}}\sin\theta_{k_z} \\
\end{array}
\right],
\end{equation}
where we introduced the notation
\begin{equation}
\cos (2\theta_{k_z})=\frac{\Delta_\mathrm{v}}{\sqrt{\Delta_\mathrm{v}^2+16t^2\cos^2(k_zc/2)}}=
\frac{\Delta_\mathrm{v}}{2E(k_z)}
\end{equation}
As a result, the eigenstates of the bulk have the form
\begin{eqnarray}
|\Uparrow\rangle_{k_z}&=&\frac{1}{\sqrt{\mathrm{M}}}\sum_{m=1}^{\mathrm{M}} e^{ik_zcm} \{\Psi_{k_z}^\uparrow|d^{(m)}_2\rangle+
\Phi_{k_z}^\uparrow|d^{(m)}_{-2}\rangle\}\otimes|\uparrow\rangle,\\
|\Downarrow\rangle_{k_z}&=&\frac{1}{\sqrt{\mathrm{M}}}\sum_{m=1}^{\mathrm{M}} e^{ik_zcm} \{\Psi_{k_z}^\downarrow|d^{(m)}_{-2}\rangle+\Phi_{k_z}^\downarrow|d^{(m)}_2\rangle\}\otimes|\downarrow\rangle.
\end{eqnarray}
They correspond to spin-up and spin-down valence band states with non-zero momentum $k_z$.
The structure of these functions defines the optical properties of the bulk.
\begin{figure}
	\includegraphics[width=8cm]{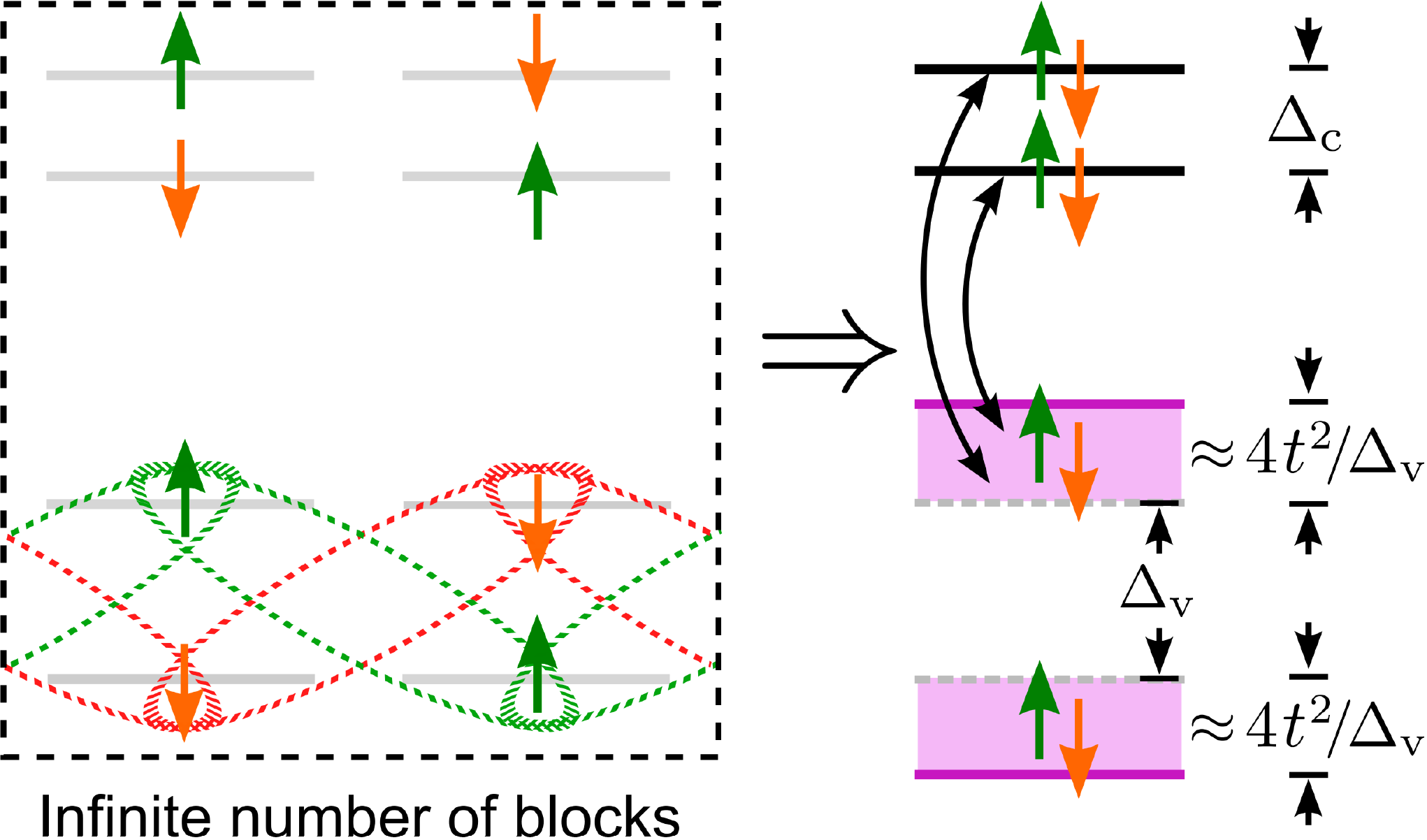}
	\caption{\label{fig:5}
		(Left panel) Diagram of the conduction and valence bands position in one of the infinite blocks, which form the bulk S-TMD. The green and red dashed ellipses represent the interaction between the states inside the block and its neighborhoods.
		(Right panel) The band structure of the bulk at the $\mathrm{K}^+$ point. Each horizontal line inside the
		purple rectangular corresponds to the state with $k_z\in[-\pi/c, \pi/c]$. The possible, lowest energy, optical transitions are presented by black solid arrow lines. The green (orange) arrows indicate the spin-up (spin-down) subbands in the conduction and valence bands.
	}
\end{figure}
Indeed, there are two subset of optical transitions from
wide valence zone (presented by the purple rectangular in Fig.~\ref{fig:5}) to spin-up and spin-down
conduction band states (presented by two straight black lines).
The first set of transitions has the energies $E_\mathrm{X_A}-E(k_z)$ with intensities
\begin{equation}
I^s_{k_z}=I_0|\Psi_{k_z}^s|^2=I_0\left(\frac{1}{2}+\frac{\Delta_\mathrm{v}}{4E(k_z)}\right).
\end{equation}
The second set of transitions has the energies $E_\mathrm{X_A}-\Delta_c-E(k_z)$ with intensities
\begin{equation}
I^s_{k_z}=I_0|\Phi_{k_z}^s|^2=I_0\left(\frac{1}{2}-\frac{\Delta_\mathrm{v}}{4E(k_z)}\right).
\end{equation}
Due to this result, the transitions between spin-up and spin-down states have the same intensity.
It reflects the absence of valley dependent circular dichroism in such system.
Moreover, the total amount of light which is absorbed by material is not changed,
but is only redistributed between the new bands.

The obtained result helps us to estimate the effective valence band mass $m_z$ in $z$ direction.
Expanding the square root at small momenta $k_z$ for high energy valence band at the $\mathrm{K}^+$ point,
we obtain $m_z=\hbar^2\Delta_\mathrm{v}/2t^2c^2$.
Substituting the previous parameters for WS$_2$ and taking  $c=12.32\mbox{{\AA}}$~\cite{schutte1987} one gets $m_z\approx3.6m_0$, where $m_0$ is free-electron mass.
Therefore the whole dispersion of the valence band of the bulk S-TMD in $\mathrm{K}^+$ point can be written as
\begin{equation}
E_\pm(\mathbf{k},k_z)= \pm\frac{\Delta_\mathrm{v}}{2} \pm \frac{4t^2}{\Delta_\mathrm{v}} -
\frac{\hbar^2\mathbf{k}^2}{2m_\pm}\mp\frac{\hbar^2k_z^2}{2m_{z}},
\end{equation}
where $\mathbf{k}=(k_x,k_y)$ are in-plane wave-vector of valence electrons.
$E_\pm(\mathbf{k},k_z)$ represent the energy dispersion of higher and lower valence bands of the bulk,
and $m_\pm$ are the effective masses of the higher and lower energy valence bands in monolayer.

\newpage
\section{Ambiguous fitting of the reflectance contrast spectrum of bilayer WS$_2$}\label{sec:rcfit:bilayer}

\begin{figure}[h]
\centering
  \includegraphics[width=1\linewidth]{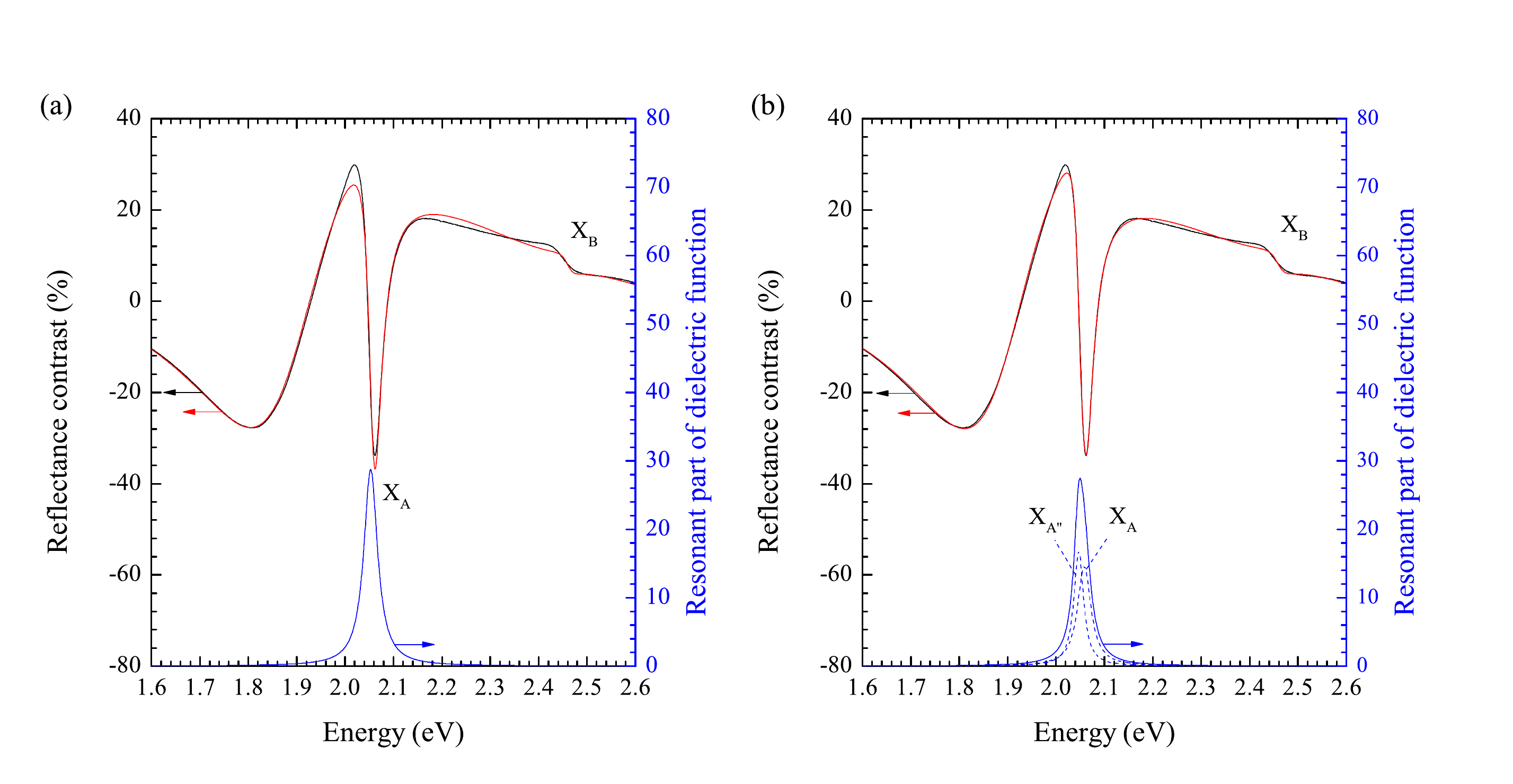}
  \vspace{-7mm}\caption{Comparison between fits (red curves) to the reflectance contrast spectrum (black curves) of bilayer WS$_2$ obtained within the framework of transfer-matrix method when assuming that the A-exciton feature in the vicinity of 2.05 eV originates from (a) single excitonic resonance (X$_{\mathrm{A}}$) and (b) two closely spaced excitonic resonances denoted by X$_{\mathrm{A}}$ and X$_{\mathrm{A"}}$. X$_{\mathrm{B}}$ stands for the B-exciton resonance. Drawn in blue are corresponding Lorentzian-type contributions to the imaginary part of dielectric function of bilayer WS$_2$ (the one associated with the B exciton is purposely not shown in the graphs).}
  \label{fig:Fit_RC_2MLs_SI}
\end{figure}

Fitting the reflectance contrast spectrum of bilayer WS$_2$ with a model curve calculated within the framework of transfer-matrix method gives an ambiguous answer to the question how many (one or two) Lorentzian-type resonances one should take into account to correctly reproduce the shape of the A-exciton feature. Solving this issue would greatly help to establish the significance of intra- and interlayer excitonic transitions described in the previous section for the absorption of light in WS$_2$ multilayers. Unfortunately, as shown in Fig.~\ref{fig:Fit_RC_2MLs_SI}, the difference between the fits obtained for one (Fig.~\ref{fig:Fit_RC_2MLs_SI}(a)) and two (Fig.~\ref{fig:Fit_RC_2MLs_SI}(b)) resonance contributions to the dielectric function of bilayer WS$_2$ in the vicinity of 2.05 eV is in our case purely quantitative. On a qualitative level, given the linewidths of the excitonic transitions involved which are as large as 35 meV, both of them equally well reproduce the experimental spectrum.  It means that in order to ultimately determine the single- or double-resonance character of the A-exciton feature in the absorption-like response of bilayer WS$_2$ one has to consider higher-quality samples with narrower excitonic transitions. As recently demonstrated for monolayers of MoS$_2$ (Ref.~\citenum{Cadiz2017}) and WSe$_2$ (Ref.~\citenum{Manca2017}), such structures can be fabricated by encapsulating the central S-TMD flake with two flakes of hexagonal boron nitride. If the reported reduction of linewidth of excitonic transitions in such samples by more than one order of magnitude with respect to the non-encapsulated ones would also hold for bilayer WS$_2$, the shape of its reflectance contrast spectrum should leave no doubt about the number of resonances contributing to the A-exction feature.

\vspace{-10mm}

\section{Simulated reflectance contrast spectra for optical transitions due to intralayer excitons in WS$_2$ multilayers}\label{sec:simrc:multilayers}

\begin{figure}[h]
\centering
  \includegraphics[width=1\linewidth]{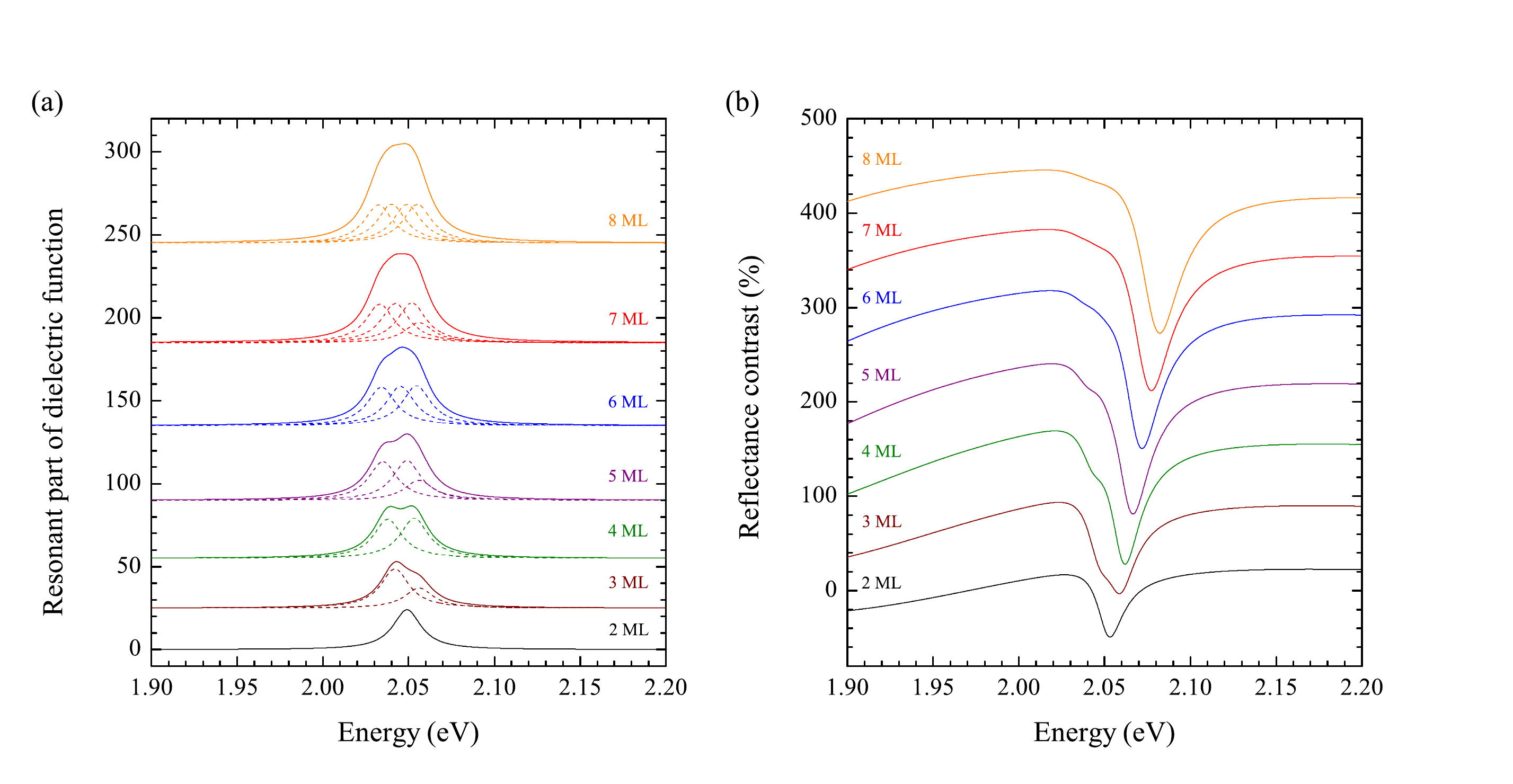}
  \vspace{-7mm}\caption{(a) Resonant contributions to the imaginary part of dielectric function of WS$_2$ multilayers and (b) corresponding reflectance contrast spectra calculated on the basis of intralayer excitonic transitions discussed in Sec.~\ref{sec:bndstrctr} and shown in Fig.~\ref{fig:2}. Drawn with dashed lines in panel (a) are individual transitions, each one with an assumed broadening of 20 meV as deduced from our experimental data.}
  \label{fig:Sim_RC_spec_AR_SI}
\end{figure}

It is already well established that the optical response of structures composed of thin films of S-TMD materials exfoliated onto alien substrates, like a piece of Si/SiO$_2$ wafer, is strongly affected by interference effects.\cite{Lien2015} Because of this fact, predicting the shape of a reflectance contrast spectrum for a set of given excitonic transitions without taking into account the internal structure of a dielectric stack under consideration represents a difficult task which may often lead to confusing and/or incorrect conclusions. It particularly applies to the intra- and interlayer excitonic transitions in WS$_2$  multilayers discussed in Sec.~\ref{sec:bndstrctr} and depicted in Fig.~\ref{fig:2}. In order to see the final shape of reflectance contrast spectra of different WS$_2$ $N$-layers supported by Si/(320\hspace{0.5mm}nm)\hspace{0.5mm}SiO$_2$ substrate, which are determined by these transitions, we first evaluated their contribution to the imaginary part of dielectric function of according WS$_2$ films by implementing them into the Lorentz-oscillator model with a broadening of every individual transition equal to 20~meV (in agreement with our experimental data). While doing so we assumed that $E_{\mathrm{X}_A} = 2.056\hspace{1mm}\mathrm{eV}$ and considered only the intralayer transitions (the upper branch of lines in Fig.~\ref{fig:2}) since the interlayer ones (the lower branch of lines in Fig.~\ref{fig:2}) with their intensities being smaller by one to two orders of magnitude than those of the intralayer transitions would have negligible impact on the resultant reflectance contrast spectra. The results of such calculations performed for WS$_2$ films composed of 2 to 8 MLs are presented in Fig.~\ref{fig:Sim_RC_spec_AR_SI}(a). In the next step, by making use of the transfer-matrix method, we simulated on their basis the corresponding reflectance contrast spectra displayed in Fig.~\ref{fig:Sim_RC_spec_AR_SI}(b). As one can see, they surprisingly well reproduce the shape and general evolution of the experimental traces shown in Fig.~11 of the main text (a small blue-shift of the feature appearing in the vicinity of 2.05 eV with increasing the number of layers may results from neglecting in our calculations the Coulomb interactions and/or the band-gap renormalization). Given the simplicity of our single-particle considerations this is quite a remarkable finding. Nevertheless, a much more detailed and accurate theory is required to confirm whether the peculiar fine structure of the A-exction line in the reflectance contrast spectra of WS$_2$ multilayers indeed originates from the intralayer excitonic transitions as discussed in Sec.~\ref{sec:bndstrctr}.

\vspace{-10mm}

\newpage
\section{Fundamental excitonic resonances as a function of temperature}\label{sec:refl:Aexciton}

\begin{center}
	\begin{figure*}[!b]
		\subfloat{}%
		\centering
		\includegraphics[width=0.33\linewidth]{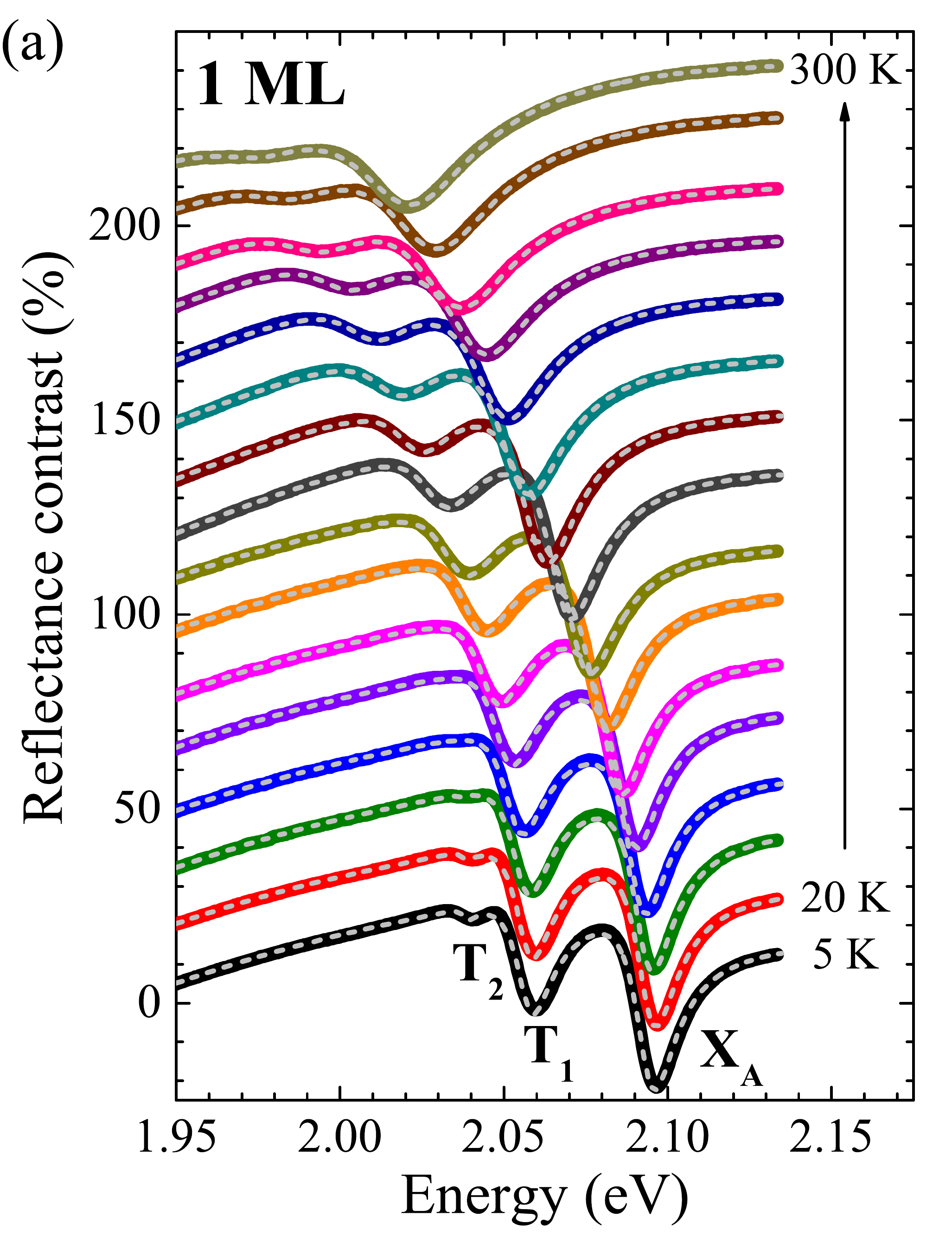}%
		\subfloat{}%
		\centering
		\includegraphics[width=0.33\linewidth]{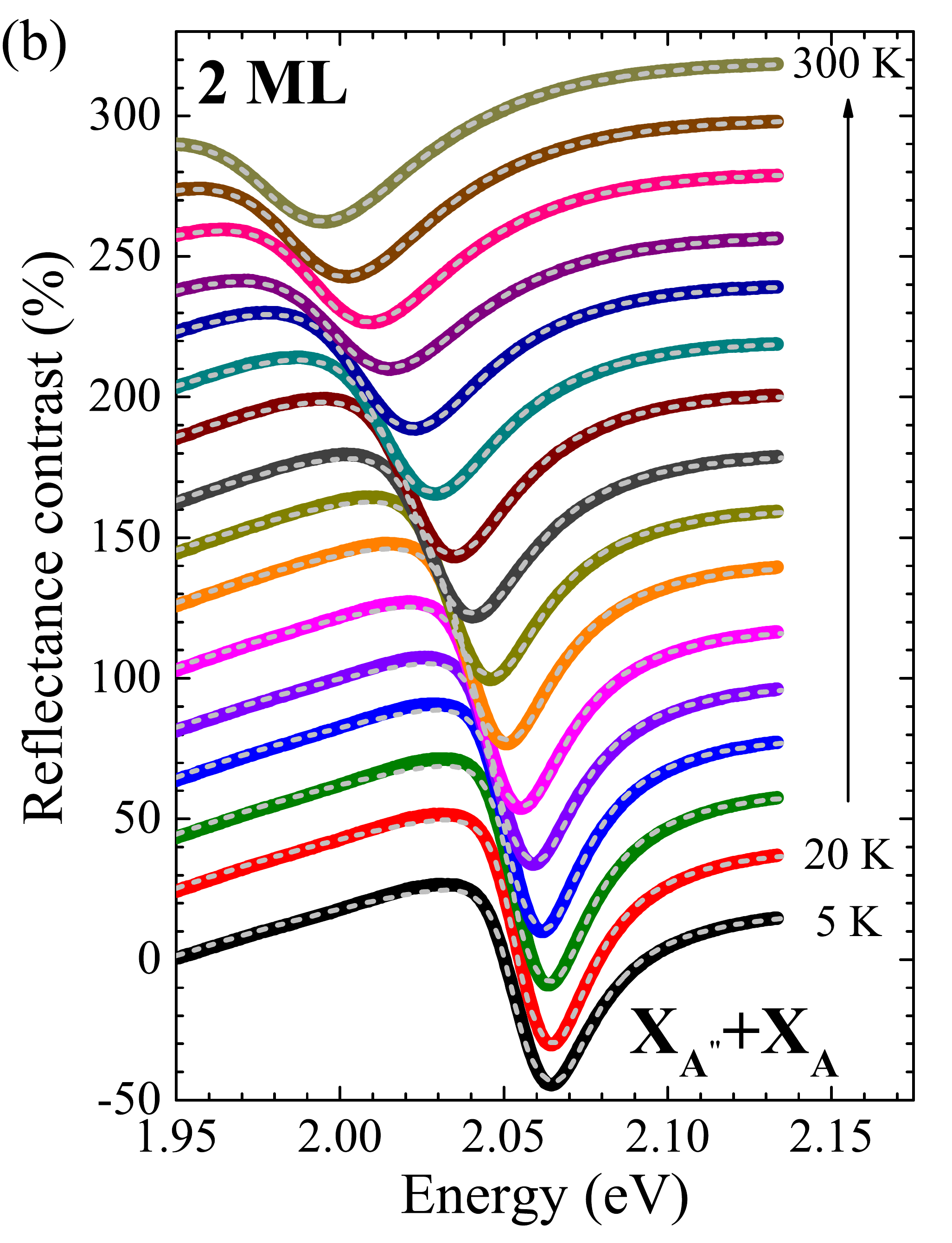}\\
		\subfloat{}%
		\centering
		\includegraphics[width=0.33\linewidth]{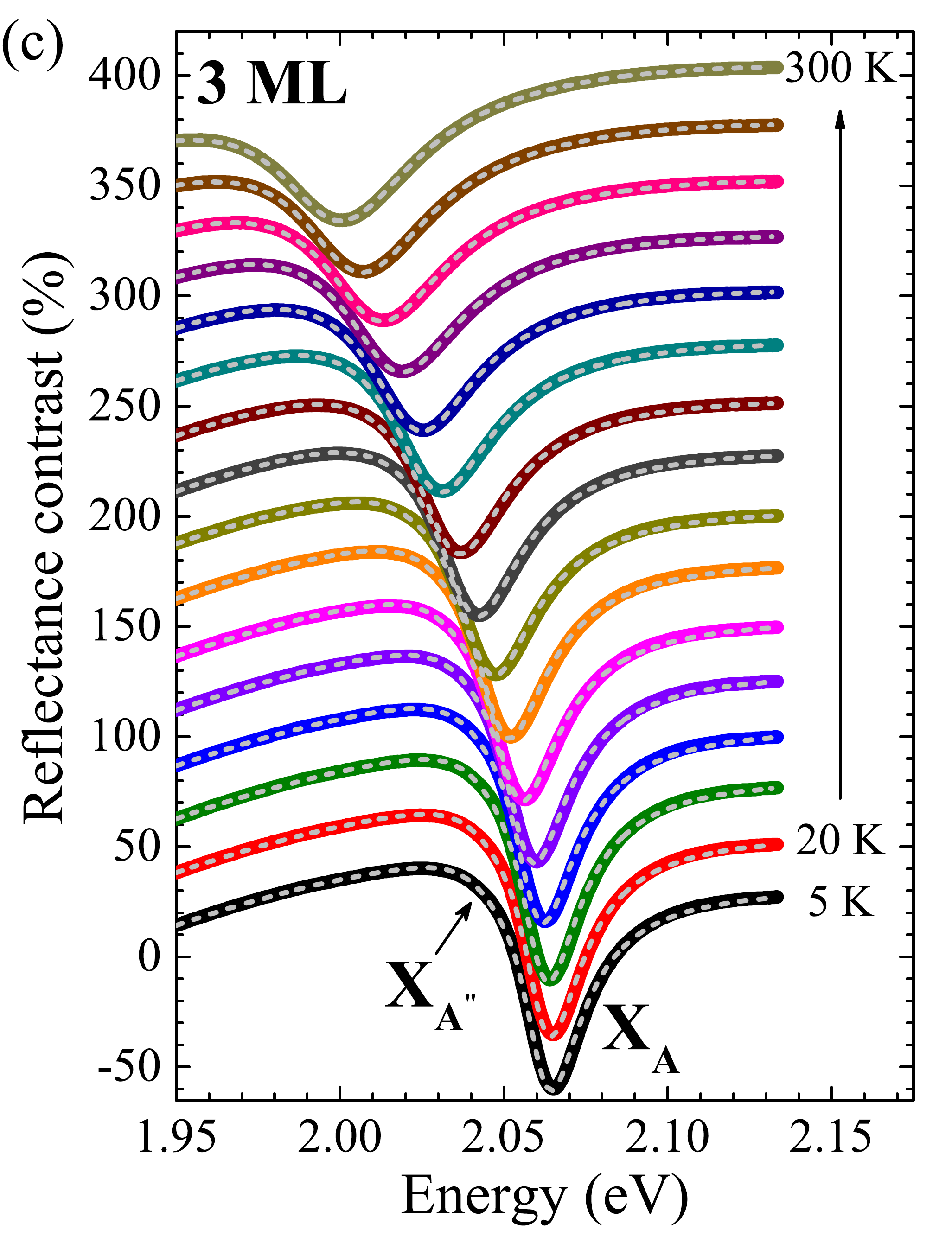}%
		\subfloat{}%
		\centering
		\includegraphics[width=0.33\linewidth]{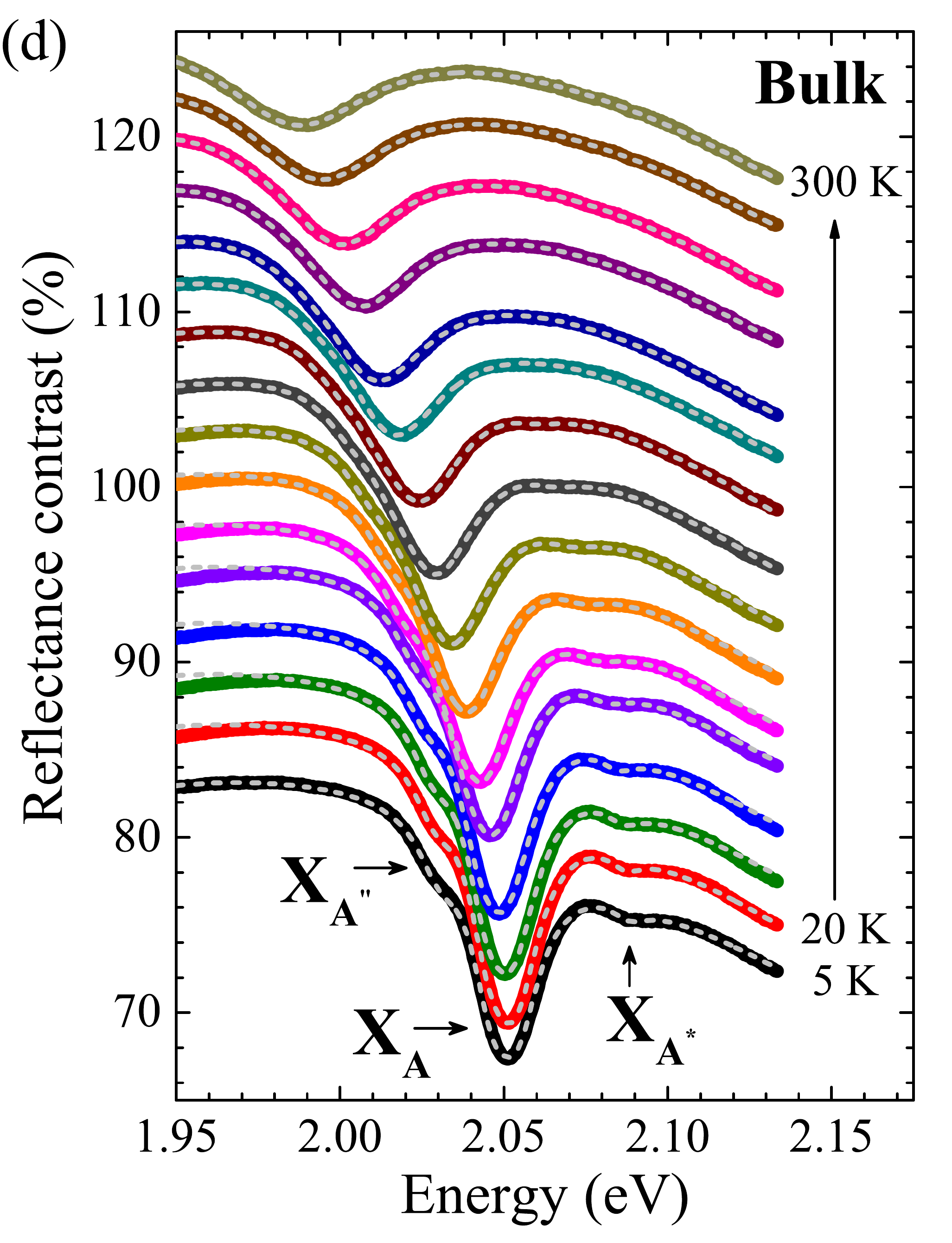}%
		\caption{RC spectra of (a) 2 ML-, (b) 3 ML-, and of the bulk flake of 32 nm thickness, measured for different temperatures ranging from 5 K up to 300 K. The dashed grey lines are the corresponding modelled curves. The spectra are vertically shifted for clarity purpose.}
		\label{fig:RvsT}
	\end{figure*}
\end{center}

Aiming at learning more about the properties of excitonic resonances associated with the transitions in the vicinity of the A exciton, we measured the RC spectra of the 1 ML-, 2 ML-, and 3 ML-thick flakes as well as of the bulk flake of 32 nm thickness as a function of temperature, see Fig. \ref{fig:RvsT}. For few-layer flakes, the temperature evolutions of the X$_\mathrm{A}$ and X$_{\mathrm{A}^"}$ transitions could be followed in the whole temperature range, whereas for the bulk flake, due to generally small changes in the RC signal even for the most pronounced features, we were not able to reliably extract the contribution coming from the X$_{\mathrm{A}^"}$ features at temperatures higher than 180~K. 

Concerning other features, the T$_2$ transition, which is clearly visible in the RC spectrum of the 1 ML at 5~K - see Fig. \ref{fig:RvsT}(a), quickly disappears from experimental traces around 40~K, which makes the analysis of its temperature evolution idle. It is not the case of an excited excitonic transition denoted by X$_{\mathrm{A}^*}$, which survives in the RC spectra of the bulk flake up to $\sim$180~K, and as such can be reliably fitted with a model dependence. For all traces presented in Fig. \ref{fig:RvsT}, the energies of all excitonic resonances experience a red shift as the temperature is being increased from 5 K up to 300 K. This type of evolution, which is characteristic of many semiconductors, can be reproduced with the aid of formulae proposed by Varshni\cite{varshni1967} and O'Donnell $et~al.$,\cite{odonnell1991} both describing the temperature dependence of the energy gap. The Varshni relation is given by:

\begin{equation}
E_g(T)=E_0-(\alpha T^2)/(T+\beta),
\label{eq:varshni}
\end{equation}

where $E_0$ stands for the band gap at absolute zero temperature, while $\alpha$ and $\beta$ represent the fitting parameters related to the electron(exciton)-phonon interaction and Debye temperature (T$_D$), respectively. From the fact that the used relation describes the temperature evolution of the band gap, it correctly reproduces also the energies of fundamental excitonic resonances, we conclude that binding energies of excitons they correspond to do not depend on temperature. 

\begin{center}
	\begin{figure}[t]
		\subfloat{}%
		\centering
		\includegraphics[width=0.33\linewidth]{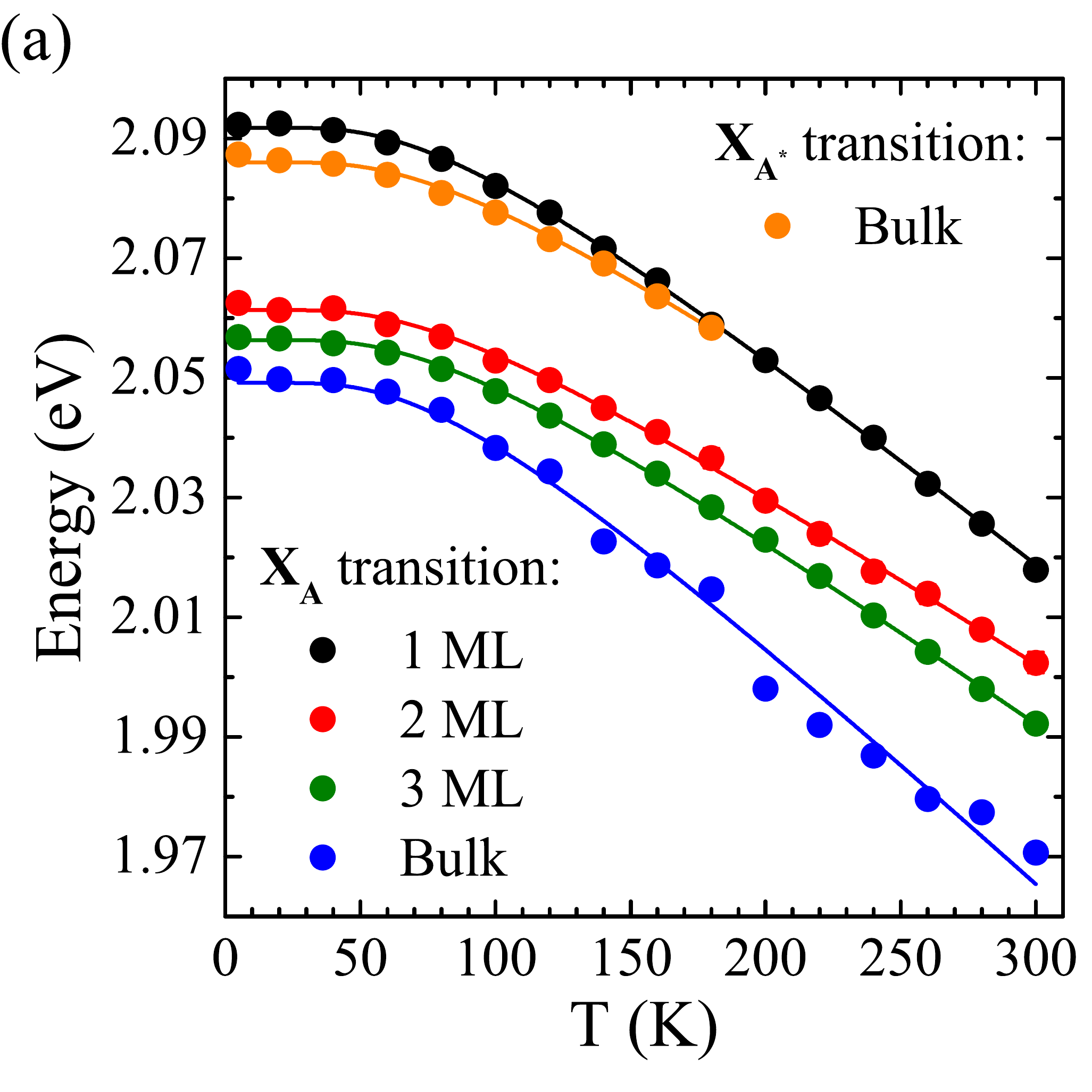}%
		\subfloat{}%
		\centering
		\includegraphics[width=0.33\linewidth]{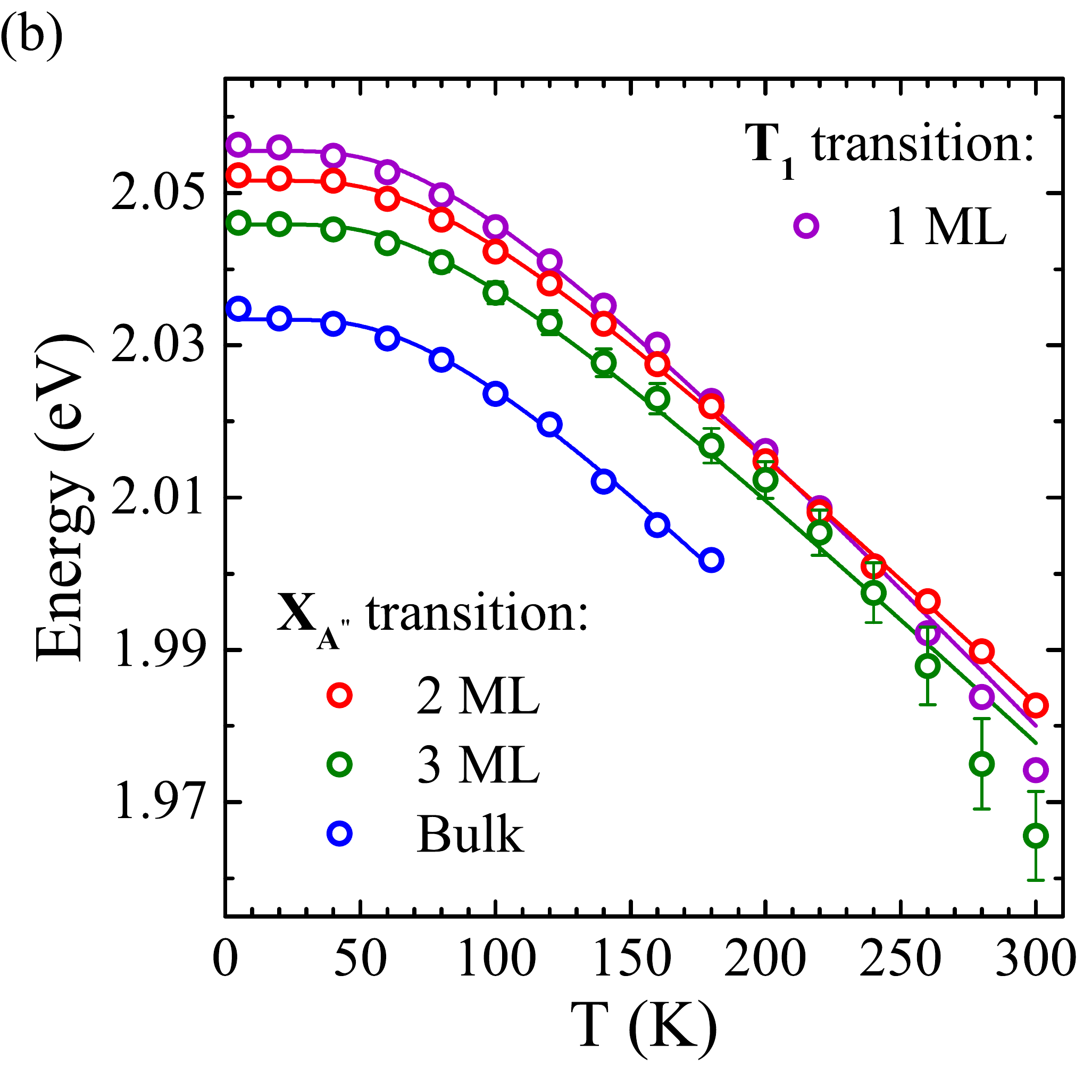}\\
		
		\subfloat{}%
		\centering
		\includegraphics[width=0.33\linewidth]{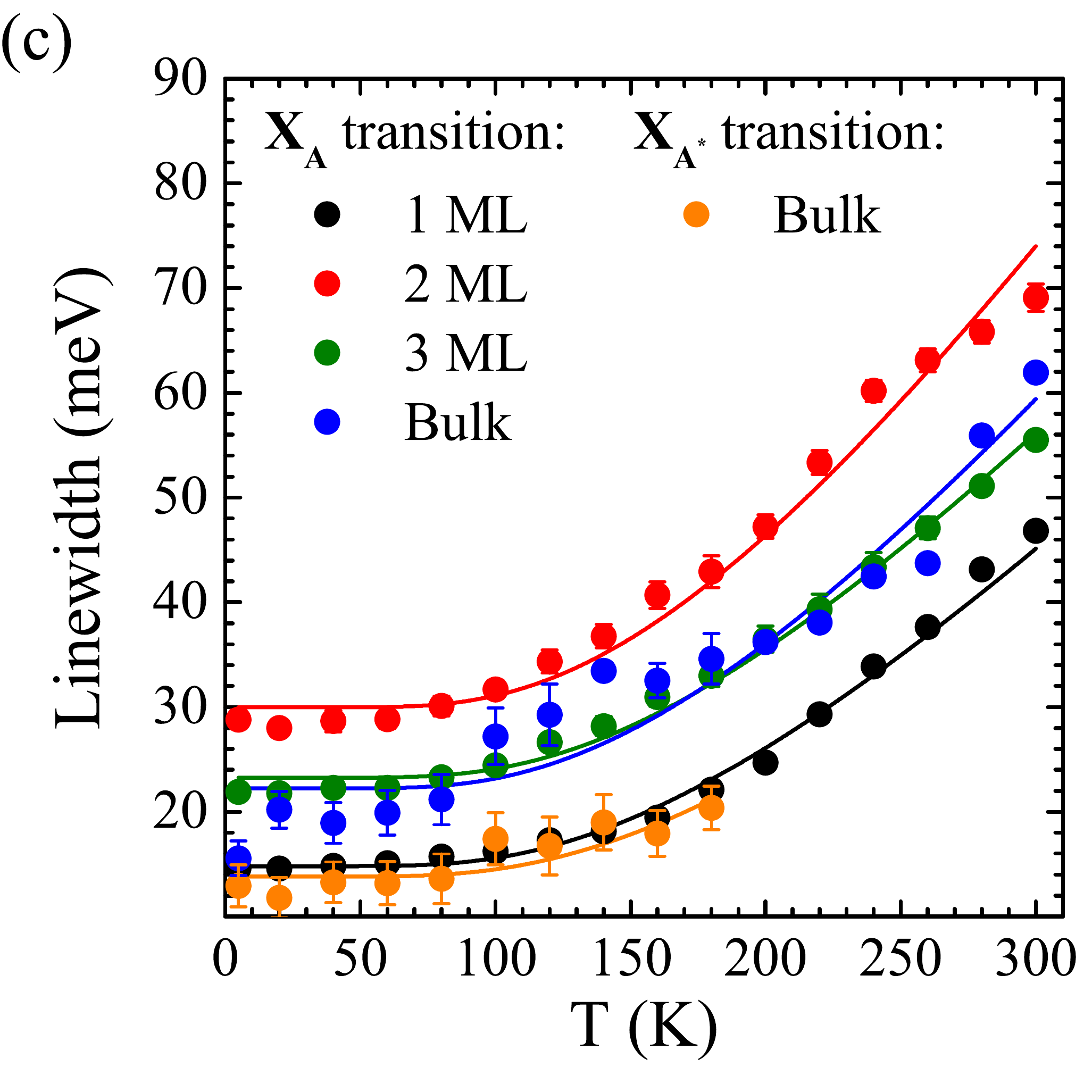}%
		\subfloat{}%
		\centering
		\includegraphics[width=0.33\linewidth]{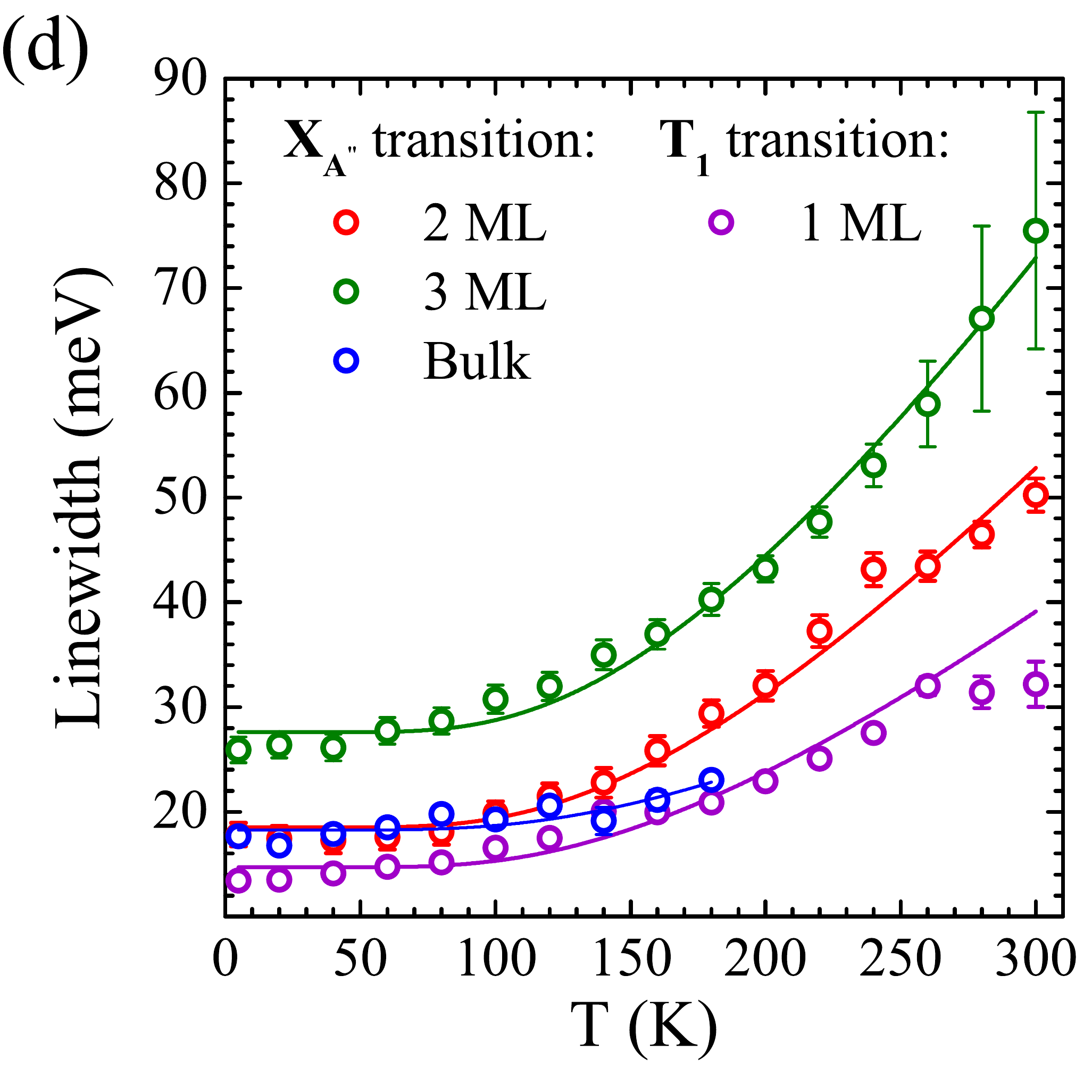}%
		\caption{Temperature evolution of the: (a) energy and (c) linewidth of the X$_{\mathrm{A}}$ andX$_{\mathrm{A}^*}$ transitions; (b) energy and (d) linewidth of the T$_1$ and X$_{\mathrm{A}^"}$ transitions extracted from the RC spectra measured on the 1 ML-, 2 ML-, and 3 ML-thick flakes, as well as on the bulk flake of 32 nm thickness. The solid circles represent the experimental results while the solid curves are fits to the data obtained with the aid of Eq. \ref{eq:odonnell} and \ref{eq:rudin}.}
		\label{fig:En_&_Width_vs_T}
	\end{figure}
\end{center}

The results of fitting our experimental data shown in Figs. \ref{fig:En_&_Width_vs_T}(a) and \ref{fig:En_&_Width_vs_T}(b) with the Varshni relation are summarized in the first row of Tab. \ref{tab:tab1}. We noticed that allowing both $\alpha$ and $\beta$ parameters to be varied at the same time led to a whole family of curves, which closely followed the experimental points but lacked the physical sense. This observation pushed us towards fixing the Debye temperature at the same level for all the flakes considered. Based on the quality of resulting fittings, we assumed that $\beta=$220~K, not far from the value reported for bulk WS$_2$ in Ref. \citenum{dumcenco2008}. In order to justify our decision, let us recall that for T $>$ T$_D$ the majority phonons that can be excited in a crystal are those with wave vectors close to the BZ boundaries. On the other hand, at low temperatures (T $\ll$ T$_D$), the only phonons that play an important role are those related to the centre of the BZ.\cite{roy2002} One may say that T$_D$ sets the lower bound for the temperature range, in which practically all possible phonon modes of a given system can be in principle observed. The number and symmetry of these modes are specified by the crystal structure, which in the case of TMDC materials does not significantly change with the layer thickness. This means that there is no reason to expect T$_D$ to be different for flakes composed of different number of layers. Concerning the value of T$_D$, as room-temperature measurements usually resolve a complete phonon spectrum of both conventional semiconductors and TMDCs, it should not be substantially larger than 300~K. Taking all these remarks into account, our choice of $\beta=$220~K dictated by the quality of fitting the experimental data with the Varshni relation seems to be a reasonable estimation. For fixed $\beta$ it can be noticed that the parameter $\alpha$ obtained for flakes of different thicknesses exhibits a larger variance for the A" transition than for the neutral excitonic resonance.

\begin{center}
	\begin{table*}
		\centering
		\caption{Parameters of fitting the temperature dependence of the resonance energy with Varshni and O'Donnell relations, and of the lineshape broadening with the Rudin relation, obtained for X$_{\mathrm{A}}$, X$_{\mathrm{A}^{*}}$, X$_{\mathrm{A}^{"}}$ and T$_1$ transitions for 1 ML, 2 ML, and 3 ML flakes and the bulk flake of 32 nm thickness.}
		\label{tab:tab1}
		\begin{tabular}{lccccccccc}
			\hline
			& \multicolumn{1}{|c }{T$_1$} & \multicolumn{3}{|c }{X$_{\mathrm{A}^"}$} & \multicolumn{4}{ |c }{X$_{\mathrm{A}}$} & \multicolumn{1}{|c }{X$_{\mathrm{A}^*}$}\\ 
			%\cline{2-5} \cline{6-9}
			Parameter & \multicolumn{1}{|c }{1 ML} & \multicolumn{1}{|c }{2 ML} & 3 ML & bulk & \multicolumn{1}{|c }{1 ML} & 2 ML & 3 ML & bulk & \multicolumn{1}{|c }{ bulk} \\
			\hline
			\multicolumn{10}{c}{Varshni's relation} \\
			\hline
			$E_0$ (eV) & \multicolumn{1}{|c }{2.058} & \multicolumn{1}{|c }{2.054} & 2.050 & 2.036 & \multicolumn{1}{|c }{2.094} & 2.063 & 2.059 & 2.053 & \multicolumn{1}{|c }{2.088}\\
			$\alpha$ (10$^{-4}$ eV/K) & \multicolumn{1}{|c }{4.32} & \multicolumn{1}{|c }{4.09} & 4.48 & 4.17 & \multicolumn{1}{|c }{4.31} & 3.52 & 3.83 & 5.02 & \multicolumn{1}{|c }{3.56}\\
			\hline
			\multicolumn{10}{c}{O'Donnell's relation} \\
			\hline
			$E_0$ (eV) & \multicolumn{1}{|c }{2.056} & \multicolumn{1}{|c }{2.051} & 2.046 & 2.033 & \multicolumn{1}{|c }{2.092} & 2.061 & 2.056 & 2.049 & \multicolumn{1}{|c }{2.086}\\
			$S$ & \multicolumn{1}{|c }{2.20} & \multicolumn{1}{|c }{2.01} & 1.99 & 2.15 & \multicolumn{1}{|c }{2.13} & 1.73 & 1.87 & 2.45 & \multicolumn{1}{|c }{1.83}\\
			\hline
			\multicolumn{10}{c}{Rudin's relation} \\
			\hline
			$\gamma_0$ (meV) & \multicolumn{1}{|c }{15} & \multicolumn{1}{|c }{19} & 23 & 18 & \multicolumn{1}{|c }{ 15} & 30 & 28 & 22 & \multicolumn{1}{|c }{14}\\
			$\gamma'$ (meV) & \multicolumn{1}{|c }{115} & \multicolumn{1}{|c }{161} & 154 & 79 & \multicolumn{1}{|c }{142} & 207 & 212 & 174 & \multicolumn{1}{|c }{126}\\
		\end{tabular}
	\end{table*}
\end{center}

The relation proposed by O'Donnell $et~al.$\cite{odonnell1991} describes the temperature dependence of the band gap in terms of an average phonon energy $<\hbar\omega>$ and reads:

\begin{equation}
E_g(T)=E_0-S<\hbar\omega>[\coth(<\hbar\omega>/2k_\mathrm{B}T)-1],
\label{eq:odonnell}
\end{equation}

where $S$ is the coupling constant and $k_\mathrm{B}$ denotes the Boltzmann constant. We found that $<\hbar\omega>$ stayed on nearly the same level, $\sim$20 meV, for all the flakes under study. As a consequence, similarly to the parameter $\beta$ discussed above, we kept it fixed during the final run of our data analysis. 

The results of fitting the experimental points with O'Donnell's relation are displayed in Figs \ref{fig:En_&_Width_vs_T}(a) and \ref{fig:En_&_Width_vs_T}(b) in the form of solid curves described by the set of parameters summarized in the second row of Tab. \ref{tab:tab1}. As can be seen, in analogy with the parameter $\alpha$ from Varshni's equation, the variance of the parameter $S$ obtained for flakes composed of different number of layers is larger for the T$_1$ and X$_{\mathrm{A}^{"}}$ features than for the X$_{\mathrm{A}}$ excitonic resonance. In spite of this similarity, it turned out that the O'Donnell relation gives significantly better fits to our data as compared to what we were able to get with the aid of Varshni's formula.

Panels (c) and (d) of Fig. \ref{fig:En_&_Width_vs_T} present the temperature evolution of the linewidth of the studied resonances, respectively. In semiconductors, such an evolution can be described by so-called Rudin's relation\cite{rudin1990} which is given by:

\begin{equation}
\gamma(T)=\gamma_0+\sigma T+\gamma'\frac{1}{(\exp^{\hbar\omega/kT}-1)},
\label{eq:rudin}
\end{equation}

where $\gamma_0$ denotes the broadening of a given spectral line at 0 K, the term linear in temperature ($\sigma$) quantifies the interaction of excitons with acoustic phonons (of negligible meaning for the present work),  $\gamma'$ arises from the interaction of excitons with LO phonons, and $\hbar\omega$ is the LO phonon energy, which was taken to be equal to 45 meV, due to the highest density of states.\cite{molina2011} Note that this value corresponds to the LO phonon energy found in an analogous experiment done on bulk WS$_2$.\cite{dumcenco2008}

The results of fitting our experimental points with Rudin's relation are presented in Figs \ref{fig:En_&_Width_vs_T}(c) and \ref{fig:En_&_Width_vs_T}(d) as solid curves described by the set of parameters shown in the third row of Tab. \ref{tab:tab1}. Due to different vertical offsets determined by $\gamma_0$, which effectively counterbalance the slopes of particular traces, they all look essentially the same concerning the rate at which the broadenings of excitonic resonances develop with increasing the temperature. One can also notice that, irrespective of the flake thickness, the values of $\gamma'$ obtained for the X$_{\mathrm{A}}$ exciton are larger than those established for the T$_1$ and X$_{\mathrm{A}^"}$ features, which indicates that phonon-assisted scattering of the former complexes is more efficient. Furthermore, in agreement with conclusions drawn in Ref. \citenum{arorawse22015} for WSe$_2$ and in Ref. \citenum{aroramose22015} for MoSe$_2$, we see a pronounced increase in $\gamma'$ between the monolayer and the bilayer for the A exciton.

\end{document}